\documentclass[review, 1p]{elsarticle}

\usepackage{lineno,hyperref}
\usepackage{amssymb}
\usepackage{amsmath}
\usepackage{tabularx}
\usepackage[english]{babel}
\usepackage{graphicx}
\usepackage{color}
\usepackage{multirow}
\usepackage{subfig} 
\usepackage{bm}
\usepackage{algorithm}
\usepackage{threeparttable}
\modulolinenumbers[5]

\journal{Journal of Computational Physics}

\begin{document}

\begin{frontmatter}

\title{A conservative sharp-interface method for compressible multi-material flows}

\author{Shucheng Pan} \ead{shucheng.pan@tum.de}
\author{Luhui Han} \ead{luhui.han@tum.de}
\author{Xiangyu Hu} \ead{xiangyu.hu@tum.de}
\author{Nikolaus. A. Adams} \ead{nikolaus.adams@tum.de}
\address{Lehrstuhl f\"{u}r Aerodynamik und Str\"{o}mungsmechanik, Technische Universit\"{a}t
M\"{u}nchen, 85748 Garching, Germany}

\begin{abstract}
In this paper we develop a conservative sharp-interface method dedicated to simulating multiple compressible fluids. Numerical treatments for a cut cell shared by more than two materials are proposed. First, we simplify the interface interaction inside such a cell with a reduced model to avoid explicit interface reconstruction and complex flux calculation. Second, conservation is strictly preserved by an efficient conservation correction procedure for the cut cell. To improve the robustness, a multi-material scale separation model is developed to consistently remove non-resolved interface scales. In addition, the multi-resolution method and local time-stepping scheme are incorporated into the proposed multi-material method to speed up the high-resolution simulations. Various numerical test cases, including the multi-material shock tube problem, inertial confinement fusion implosion, triple-point shock interaction and shock interaction with multi-material bubbles, show that the method is suitable for a wide range of complex compressible multi-material flows.
\end{abstract}

\begin{keyword}
compressible multi-material flows, sharp interface method, multi-resolution simulations, level-set method, interface scale separation
\end{keyword}

\end{frontmatter}


\section{\label{sec:intro}Introduction}

The compressible multi-material problems occur in a broad range of scientific and engineering areas such as high energy physics and astrophysics. Typical example includes inertial confinement fusion (ICF) \cite{thomas2012drive}, core-collapse supernova \cite{lentz2015three} and hypervelocity impact \cite{gisler2003two}. In these problems, different materials separated by the interface have significantly different material properties and equation of states (EOS). Large density or pressure jumps inside the material or across the interface may occur and leads to complicated flow fields and interface evolution. For these compressible multi-material problems, numerical modeling has received increasing attention in recent years due to its lower cost and higher flexibility than experimental investigation. Many well-established methods have been proposed to simulate two-phase compressible flows, such as front-tracking method \cite{brakke1992surface}, arbitrary-Lagrangian-Eulerian (ALE) method \cite{hirt1974arbitrary}, volume-of-fluid (VOF) \cite{hirt1981volume} and level-set method \cite{osher1988fronts}. Among these, sharp interface method has the advantage of introducing no additional nonphysical mixing in interface treatment during simulations of immiscible materials or shock-driven miscible materials.

Unlike the front-tracking and ALE methods, VOF and level-set methods can automatically avoid the difficulty in capturing complex geometries of interfaces as it implicitly defines the interface and solves its evolution. Although VOF inherently preserves conservation, it treats the interface with a smeared interface representation which is not suitable for immiscible materials problems or extremely fast high energy processes. And when more than two materials involved, complicated interface reconstruction algorithms, such as onion-skin \cite{benson2002volume} and serial-dissection \cite{dyadechko2008reconstruction}, are used to find interface locations from the volume-faction data, usually relying on a trivial material ordering strategy. The mixed treatment for fluid states inside a cell occupied by more than two materials (hereafter referred to as multi-material-cell) in the VOF method does not ensure a sharp-interface property \cite{so2012anti}. In the level-set method, the interface reconstruction is straightforward via a signed distance function as the interface implicitly represented by the zero contour which can be considered as a non-smeared interface representation. The sharp-interface property can be imposed by interface-interaction treatment \cite{hu2004interface} for two-phase flows. For multi-material flows involving more than two fluids, difficulties arise when the interface-network motion is captured and the interface-interaction inside a multi-material-cell is performed. The first issue can be addressed by a recently developed multi-region level-set method \cite{pan2017high} which outperform the multiple level-set method \cite{starinshak2014new} on efficiency and the regional level-set method \cite{zheng2009simulation} on accuracy. The latter one can be handled by applying the ghost fluid method \cite{fedkiw1999non} on the multiple level-set functions to avoid complex interface interaction, however, it leads to the nonconservation issue \cite{wang2009improved, starinshak2014new}. Interface interaction model with conservation-preserving property has been developed for two-phase flows \cite{olsson2005conservative, hu2006conservative}, but, to our knowledge, not been proposed for more than two fluids.

Numerical methods may suffer from a lack of robustness when complex interface topology changes, such as sudden generation and destruction of small thin filaments and isolated droplets, are encountered. Interface scale separation models based on refined level-set grid method \cite{herrmann2008balanced, herrmann2010parallel} and identifying resolved/non-resolved interface segments \cite{hu2010multi} have been proposed for two-phase flows to remove non-resolved structures. More recently a model empolying the constrained stimulus-response procedure \cite{han2015scale} is developed for interface scale separation to increase the robustness for simulations of compressible interfacial flows, whose computational efficiency has been improved \cite{luo2016efficient}. However, the scale separation model for more than two materials is not developed in the literatures and the extension of previous models \cite{herrmann2008balanced, herrmann2010parallel, han2015scale, luo2016efficient} is not straightforward.

The objective of the present paper is to develop an efficient and robust numerical method for compressible multi-material flows. In order to ensure conservation and sharp-interface property, several operations related to the interface network are proposed. First of all, the interface-network evolution is captured accurately and efficiently by a recently developed multi-region level-set method which is a combination of original level-set and regional level-set methods to adopt the respective advantages of these two methods. In order to maintain conservation and impose sharp-interface treatment, we extend the two-phase conservative sharp-interface method by introducing a conservation correction and a reduced interface-interaction model in each multi-material-cell. 
A multi-material interface scale separation model is proposed to remove non-resolved interface segments and thus increase the robustness in high-resolution simulations. The paper is organized as follows. In Sec. \ref{sec:method}, our multi-material sharp-interface method, including the multi-region level-set method, conservative finite volume method, reduced interface-interface model and multi-material interface scale separation operation is detailed. The proposed method gains computational efficiency by the multi-resolution method and local time-stepping scheme. The accuracy, capability and robustness of the method are demonstrated in Sec. \ref{sec:validation} by a range of numerical examples, followed by a brief conclusion in Sec. \ref{sec:conclusion}.
\section{Numerical method \label{sec:method}}
\subsection{Governing equations}
The governing equations of invisicid compressible flows are
\begin{equation}
\frac{\partial \mathbf{U}}{\partial t}+\nabla \cdot \mathbf{F}(\mathbf{U})=0,
\end{equation}
where $\mathbf{U} = (\rho$, $\rho u$, $\rho v$, $\rho w$, $\rho E)^T$, in which $\rho$, $u$, $v$, $\rho w$, and $\rho E$ are the density, the three velocity components and the total energy with relation $E = e + \frac{1}{2}(u^2 + v^2 + w^2)$, with $e$ being the internal energy per unit mass. The inviscous flux tensor $\mathbf{F}$ is
\begin{equation}\label{govern}
\mathbf{F}(\mathbf{U})=
\left[\begin{array}{ccc}
    \rho u& \rho v & \rho w \\
    \rho u^2+p& \rho vu & \rho wu \\
    \rho uv & \rho v^2+p & \rho wv \\
    \rho uw & \rho vw & \rho w^2+p \\
    u(\rho E+p) & v(\rho E+p) & w(\rho E+p) \\
\end{array}\right].
\end{equation}
To close the governing equations EOS is required to describe the thermodynamic properties of the materials. The EOS for an idea gas states
\begin{equation}
p=(\gamma-1)\rho,
\end{equation}
where $\gamma$ is the ratio of specific heats. While for water-like fluids, the pressure is determined by Tait's equation of state
\begin{equation}
p=B(\frac{\rho}{\rho_0})^\gamma-B+p_0,
\end{equation}
where $\rho_0$ and $p_0$ are the reference density and pressure, and $B$ is a constant. Other EOSs may also be employed in compressible multi-material flows, such as the stiffened gas EOS for water under very high pressure, the Jones-Wilkins-Lee EOS for the detonation-products gas and the Mie-Gr\"{u}neisen EOS for solid mechanics \cite{hu2009hllc}.

The multi-material fluid dynamic problem contains multiple different fluids. Assuming there are $\mathcal{N}$ materials (or fluids) in the problem to partition the entire domain, $\Omega = \bigcup_{\chi \in \mathrm{X}} \Omega^{\chi}$, where $\Omega^{\chi}$ is the material domain and $\mathrm{X}=\{\chi \in \mathbb{N} | 1\leq \chi \leq \mathcal{N} \}$ is the index set for all materials. Given $\xi, \eta \in \mathrm{X}$ and $\xi \neq \eta$, we define $\partial \Omega^{\chi}$ as the material boundary, $\Gamma_{\xi \eta} = \partial \Omega^{\xi} \bigcap \partial \Omega^{\eta} $ as the pairwise material interface that separates the two material domains, $\Gamma = \bigcup \Gamma_{\xi \eta}$ as the interface network, and $\mathrm{J}=\bigcap \Gamma_{\xi \eta}$ as the multiple junctions. Each material has its individual material parameters (such as viscosity and surface tension) and EOS.
\subsection{Interface capturing \label{multiregion-level-set}}
To capture the evolution of complex interface networks, we have developed the multi-region level-set method which is a combination of the original level-set method \cite{osher1988fronts} and the regional level-set method \cite{zheng2009simulation}. Globally we use a single regional level-set for representing the multi-region (or multi-material in this paper) system to significantly reduce the memory cost. Locally we construct multiple local level-set functions to directly capture the evolution of the interface network in order to save computational effort and avoid inaccuracies in reconstructing the interface network. Although the total number of materials may be very large, locally the number of materials is limited, so solving multiple locally constructed level-set advection equations is efficient. This method can achieve high-order accuracy for pure advection and rotation cases without artifacts generation \cite{pan2017high}.

The total system can be represented by regional level-set function $\varphi^{\chi}(\mathbf{x})=(\varphi(\mathbf{x}),\chi(\mathbf{x}))$, where $\varphi(\mathbf{x})\geq 0$ is the unsigned distance function and $\chi(\mathbf{x})$ is an integer material indicator. Then the material domain $\Omega^{a}$ is identified by the indicator, $\Omega^{\xi} = \{\mathbf{x}|\chi(\mathbf{x})=\xi\}$, and the interface network is defined as $\Gamma = \{\mathbf{x}|\varphi(\mathbf{x})=0\}$. On a two-dimensional uniform Cartesian grid, regional level-set $\varphi^{\chi}_{i,j}=(\varphi_{i,j},\chi_{i,j})$ is defined at the center of the finite-volume cell $C_{i,j}$. The $\mathcal{N}_s$ materials contained in a local set of cells $V_s=\{C_{k,l} | i-1<k<i+1, j-1<l<j+1\}$ are labelled indirectly by a local index set $\mathrm{X}_s= \{r \in \mathbb{N} | 1\leq r \leq \mathcal{N}_s \}$.

By using a construction operator for generating the local multiple level-set fields and a reconstruction operator for reconstructing the global regional level-set field from the local level-set fields, the evolution step contains three main procedures for $C_{i,j}$:\\
(1) Construct $\mathcal{N}_s$ local level-set fields $\phi^{r,n}_{k,l}$ for the current time-step $n$ at the center of each cell $C_{k,l}$ which belongs to the spatial discretization stencil of $C_{i,j}$:
\begin{equation}\label{construction}
\phi^{r,n}_{k,l}= \mathbf{C}_r \left(\varphi^{\chi}_{k,l}\right) =
\begin{cases}
\varphi_{k,l} \quad {\rm if}~ \chi_{k,l} = \chi_r \cr
-\varphi_{k,l} \quad \rm{otherwise}	
\end{cases}, \quad r \in \mathrm{X}_s
\end{equation}
(2) Compute new $\phi^{r,n+1}_{i,j}$ at the next time-step $n+1$ by solving the $\mathcal{N}_s$ local advection equations. The formulation of local advection equation depends on $\mathcal{N}_s$. If $\mathcal{N}_s\leq 2$ it recovers the original level-set advection equation while if $\mathcal{N}_s \geq 3$ it can be rewritten as
\begin{eqnarray} \label{local-advection}
&\phi^{r, (s+1)}_{i,j} = \beta_s\phi^{r, n}_{i,j}
+ (1-\beta_s)\left[\phi^{r, (s)}_{i,j} -
\Delta t \mathbf{v}^{n}_{i,j}\cdot\left(\nabla \phi^{r}\right)^{(s)}_{i,j}\right],\quad 0 \leq s \leq  m, r \in \mathrm{X}_s,
\nonumber \\
&\phi^{r, (0)}_{i,j} = \phi^{r, n}_{i,j}, \quad \phi^{r, (m+1)}_{i,j} = \phi^{r, n+1}_{i,j},
\end{eqnarray}
where $m$ is the number of Runge-Kutta sub-steps, $\beta_s$ is the parameter in the $s$-th sub-step, $\mathbf{v}$ is the advection velocity, and $\left(\nabla \phi^{r}\right)^{(s)}_{i,j}$
is the finite difference approximation of the spatial derivative at the center of a finite-volume cell $C_{i,j}$.\\
(3) Reconstruct the new regional level-set $\varphi^{\chi}_{i,j}$ at the center of $C_{i,j}$ from the $\mathcal{N}_s$ new local level-set fields $\phi^{r,n+1}_{i,j}$ by the reconstruction operator $\mathbf{R}$
\begin{equation}\label{reconstruction}
\varphi^{\chi}_{i,j} = \mathbf{R} \left(\phi^{r,n+1}_{i,j}, r \in \mathrm{X}_s\right)
=  \left(\left|\max \phi^{r,n+1}_{i,j}\right|, \arg\max_{\chi_r} \phi^{r,n+1}_{i,j}\right).
\end{equation}
\subsection{Conservative sharp-interface method\label{conservative-method}}
Here we extend the two-phase discretized governing equation of conservative sharp-interface method \cite{hu2006conservative} to multi-material flows. In a two-dimensional uniform Cartesian grid with grid spacings $\Delta x$ and $\Delta y$, the flow variable $\mathbf{U}$ is defined at the center of the finite-volume cell. For each material $\chi$ residing in cell $C_{i,j}$, we can integrate Eq. \ref{govern} over the space-time volume $C_{i,j}\cap \Omega^{\chi}(t)$ and apply divergence theorem to obtain
\begin{equation}
\int_{n}^{n+1} \mathrm{d}t \int_{\alpha_{i,j}^{\chi}(t)} \mathrm{d}x \mathrm{d}y \; \frac{\partial \mathbf{U}}{\partial t} + \int_{n}^{n+1} \mathrm{d}t \int_{\partial C_{i,j} \cap \Omega^{\chi}(t)} \mathrm{d}x \mathrm{d}y \; \mathbf{F} \cdot \mathbf{n}= 0,
\end{equation}
where $\alpha_{i,j}^{\chi}(t)$ is the time dependent volume fraction of material $\chi$ in $C_{i,j}$. $\partial C_{i,j} \cap \Omega^{\chi}(t)$ contains two parts: one is the combination of the four segments of the cell faces after cut by the material interface, which can be written in the form of $A_{i+1/2,j}^{\chi}(t) \Delta y$, $A_{i,j+1/2}^{\chi}(t) \Delta x$, $A_{i-1/2,j}^{\chi}(t) \Delta y$, and $A_{i,j-1/2}^{\chi}(t) \Delta x$, where $A^{\chi}(t)$ is the aperture, see Fig. \ref{stencil}(a); the other one, denoted as $\Delta \Omega^{\chi}_{i,j}$, is the segment of material boundary $\partial \Omega^{\chi}$ inside the cell $C_{i,j}$.
This integral equation can be discretized with explicit Euler time marching scheme as
\begin{eqnarray}\label{discretized_equ}
&&\alpha_{i,j}^{\chi,n+1}\mathbf{U}_{i,j}^{n+1}=\alpha_{i,j}^{\chi,n}\mathbf{U}_{i,j}^{n}+ \frac{\Delta t}{\Delta x} \left[A_{i-1/2,j}^{\chi}\mathbf{\hat{F}}_{i-1/2,j} - A_{i+1/2,j}^{\chi}\mathbf{\hat{F}}_{i+1/2,j}\right] \nonumber \\
&& + \frac{\Delta t}{\Delta x} \left[A_{i,j-1/2}^{\chi}\mathbf{\hat{F}}_{i,j-1/2} - A_{i,j+1/2}^{\chi}\mathbf{\hat{F}}_{i,j+1/2}\right] + \frac{\Delta t}{\Delta x \Delta y}\mathbf{\hat{X}}(\Delta \Omega^{\chi}_{i,j}),
\end{eqnarray}
where $\Delta t$ is the time step size determined by CFL condition. $\alpha_{i,j}^{\chi}\mathbf{U}_{i,j}$ is the conservative quantity vector in $C_{i,j}$, with $\mathbf{U}_{i,j}$ being the cell-averaged quantity vector of the material $\chi$. $\mathbf{\hat{F}}$ is the numerical flux at cell-face and $\mathbf{\hat{X}}(\Delta \Omega^{\chi}_{i,j})$ is the momentum and energy exchange flux determined by the interface interaction model discussed in Sec. \ref{separation-model}.
\subsubsection{Conservation correction for a multi-material-cell\label{conservaive-correct}}
\begin{figure}[p]
\begin{center}
\includegraphics[width=0.8\textwidth]{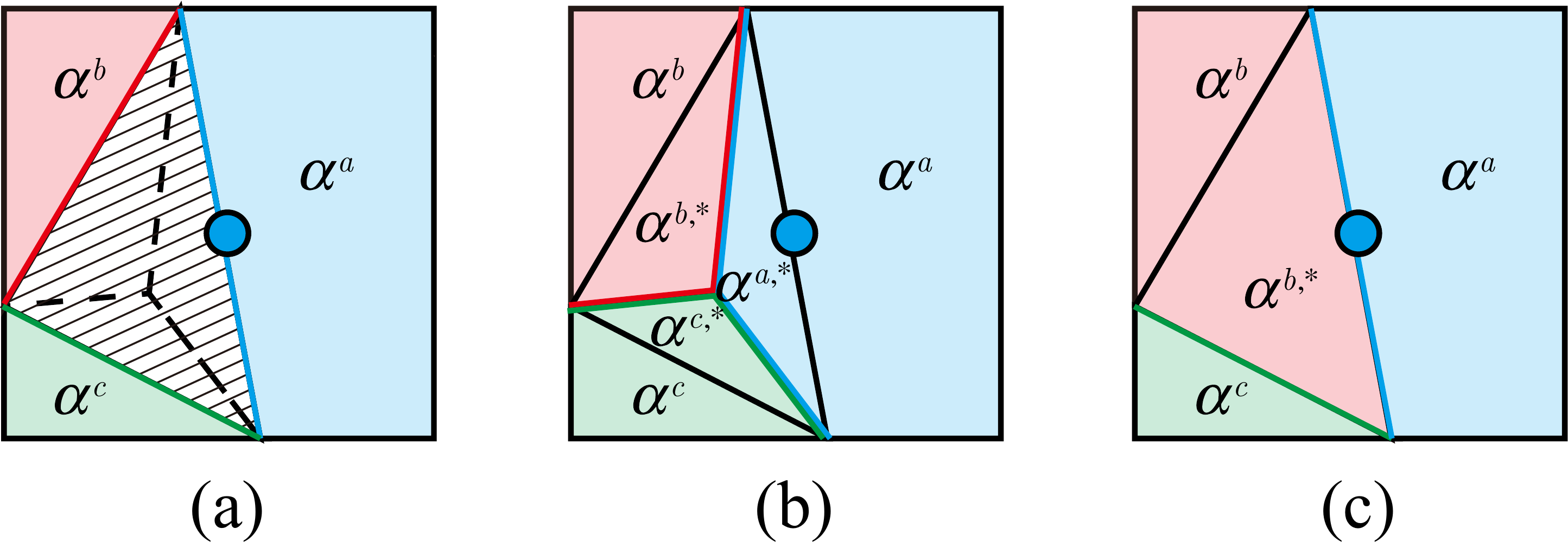}
\caption{A schematic representations of volume fraction correction for a multi-material finite-volume cell to maintain conservation: (a) without modification, (b) sub-cell reconstruction and (c) conservation modification.}
\label{subcell_volume}
\end{center}
\end{figure}

For the material $\chi$ summing Eq. (\ref{discretized_equ}) over the its material domain $\Omega^{\chi}$ yeilds
\begin{equation}\label{sum_equ}
\sum_{i,j} \alpha_{i,j}^{\chi,n+1}\mathbf{U}_{i,j}^{n+1} = \sum_{i,j} \alpha_{i,j}^{\chi,n}\mathbf{U}_{i,j}^{n}+ \sum_{i,j} \frac{\Delta t}{\Delta x \Delta y}\mathbf{\hat{X}}(\Delta \Omega^{\chi}_{i,j}) + \mathrm{boundary} \ \mathrm{terms}
\end{equation}
For two-phase flows, overall conservation can be achieved by summing Eq. (\ref{sum_equ}) for the two fluids because interface-exchange term in cut cell always has opposite sign and the sum of volume fractions of the two fluids equals $1$. While for multi-material flows, although the volume fractions in a two-material-cell satisfies $\alpha^{\chi_1}+\alpha^{\chi_2}=1$, in a multi-material-cell where more than two materials meet the sum of volume fractions may be $\sum_{r \in \mathrm{X}_s} \alpha^{\chi_r} \neq 1.0$ if we assume the material boundary segment is piecewise planar inside a cell. For example in Fig. \ref{subcell_volume}(a), the planar material boundary segments of tree materials never coincide, resulting a shadowed void region, and thus $\alpha^{a}+\alpha^{b}+\alpha^{c}<1.0$. To correct this inaccurate total volume fraction, the piecewise planar assumption no longer holds true in this cell. A straightforward way is to explicitly reconstruct the sub-cell topology inside this multi-material-cell as shown in Fig. \ref{subcell_volume}(b). Note that all the three material boundary segments are not planar now. The new volume fraction for each material is upadted by
\begin{equation}\label{vol_corr1}
\alpha^{a} = \alpha^{a}+\alpha^{a,*}, \ \alpha^{b} = \alpha^{b}+\alpha^{b,*}, \ \alpha^{c} = \alpha^{c}+\alpha^{c,*}.
\end{equation}
Although this procedure is considered to be accurate as long as the explicit reconstruction method is sufficiently accurate, it is computational expensive especially in three dimensions. An alternative efficient way plotted in Fig. \ref{subcell_volume}(c) suggests that instead of explicitly reconstructing the sub-cell structure, we only modify the volume fraction corresponding to the least mass inside this cell:
\begin{equation}\label{vol_corr2}
\alpha^{\chi^*} = \alpha^{\chi^*}+\alpha^{\chi^*,*}, \quad  \alpha^{\chi^*,*} = 1- \sum \alpha^{\chi}, \quad \chi^* = \arg\min_{\chi} (\alpha^{\chi} \rho_{\chi}).
\end{equation}
Thus this method is efficient especially in 3D as no reconstruction is needed. Although it does not resolve the sub-cell structure and is considered less accurate than the first method, the error is limited as the dominated material (labelled by primary indicator) inside this cell has a volume fraction larger than $0.5$ and it does not change its value during correction.
\subsubsection{Material interface interaction model\label{interaction-model}}
\begin{figure}[p]
\begin{center}
\includegraphics[width=0.95\textwidth]{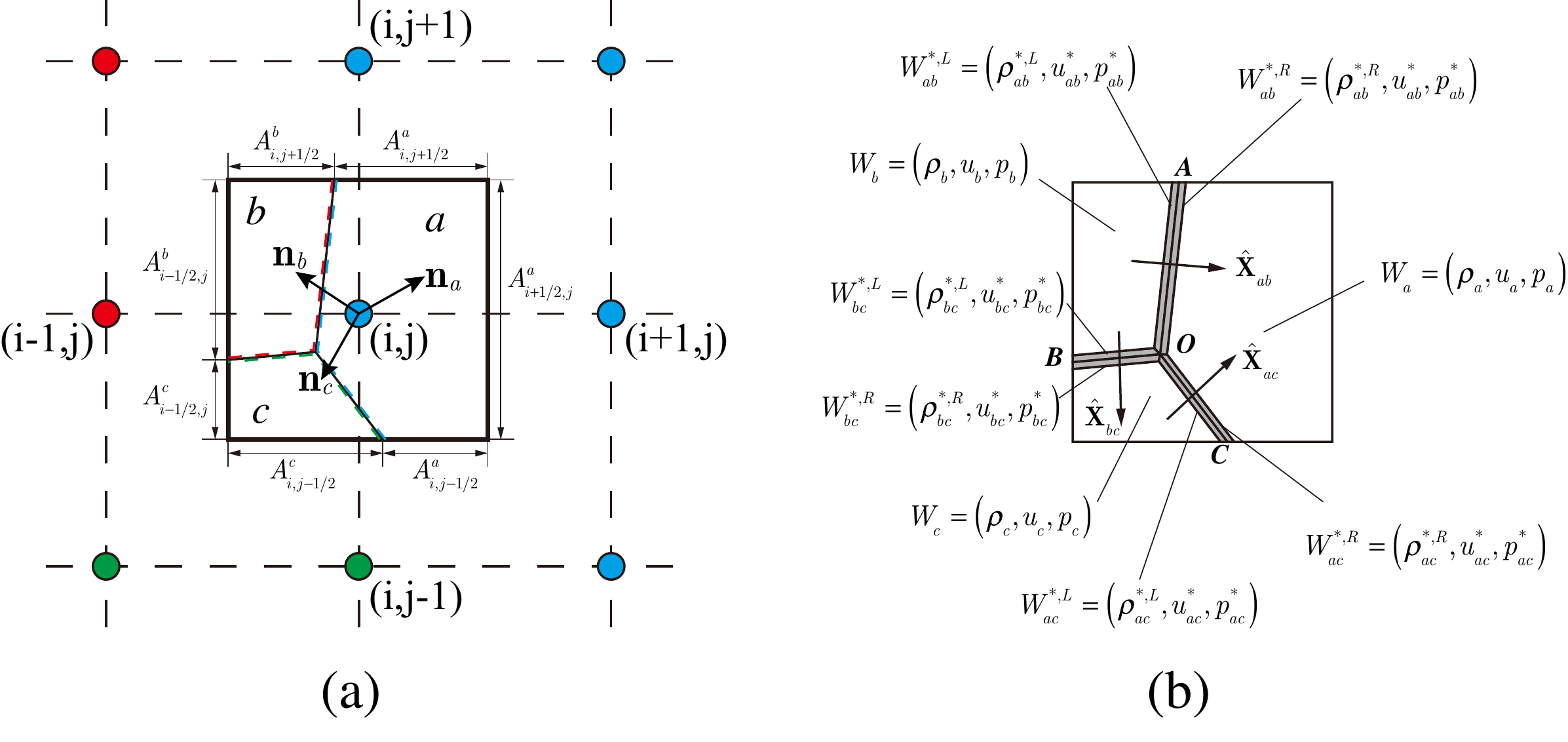}
\caption{Schematic of conservative discretization for a 3-material cell.}
\label{stencil}
\end{center}
\end{figure}

For each connected material pair $(\Omega^{\xi}, \Omega^{\eta})$, its interface interaction occurs at the pairwise interface $\Gamma_{\xi \eta}$ and is described by a Riemann problem.
Solving this Riemann problem obtains the interface condition which is used to calculate the exchange flux $\mathbf{\hat{X}}$ in Eq. (\ref{discretized_equ}) and the interface advection velocity $\mathbf{v}$ in Eq. (\ref{local-advection}). Hu et. al \cite{hu2009hllc} has proved that the HLLC Riemann solver \cite{toro1994restoration} is roust, accurate and efficient to handle two-phase flow with very strong interaction and large jumps of material properties. In this paper, we employ their implementation to solve the multi-material interface interactions. For instance, the finite-volume cell $C_{i,j}$ in Fig. \ref{stencil}(a) is occupied by three materials $a$, $b$ and $c$ which are colored by blue, red and green, respectively. According to Fig. \ref{stencil}(b), the interface condition at each pairwise interface segment $\Gamma_{\xi \eta}$ is obtained by solving a corresponding 1D Riemann problem $\mathsf{R}(W_L, W_R)$ along a local coordinate $\mathbf{n}_{\xi \eta}$, i.e., the normal direction of $\Gamma_{\xi \eta}$. Then the constant left and right states in the Riemann problem, $(W_L, W_R)$, are defined as
\begin{equation}
(W_L, W_R) = \begin{cases}
(\hat{W_{\xi}}, \hat{W_{\eta}}) \quad {\rm if}~ \alpha^{\xi} \rho_{\xi}  < \alpha^{\eta} \rho_{\eta}  \cr
(\hat{W_{\eta}}, \hat{W_{\xi}}) \quad \rm{otherwise}
\end{cases}, \hat{W_{\xi}} =
\begin{cases}
W_{\xi} \quad {\rm if}~ \xi = \chi_1  \cr
W_{\xi}^g \quad \rm{otherwise}
\end{cases},
\end{equation}
where the superscript ``g'' indicates the ghost state at the cell center.
Then we rewrite the EOS with a general form
\begin{equation}
p = p(\rho, e),
\end{equation}
and the speed of sound $c_s$ with
\begin{equation}
c_s = \left. \frac{\partial p}{\partial \rho} \right|_e + \left. \frac{p}{\rho^2} \frac{\partial p}{\partial e} \right|_{\rho} = \Psi + \Upsilon \frac{p}{\rho},
\end{equation}
where $\Upsilon$ is the Gr\"{u}neisen coefficient and $\Psi$ determines the material properties \cite{hu2009hllc}. Afterwards, the HLLC solver is invoked to obtain the two intermediate states, $W^{*,L}_{\xi \eta}$, $W^{*,R}_{\xi \eta}$ (see Fig. \ref{stencil}(b)) which have the relation
\begin{equation}
u^{*,L}_{\xi \eta} = u^{*,R}_{\xi \eta} = S_M = u^*, \quad p^{*,L}_{\xi \eta} = p^{*,R}_{\xi \eta} = p^*,
\end{equation}
where $S_M$ is the speed of contact wave or material interface. Toro et al. \cite{toro1994restoration} use the jump conditions and the integral form of conservation law to obtain the normal contact wave velocity
\begin{equation}
u^* = S_M = \frac{(S_R-u_R)\rho_R u_R + (u_L-S_L)\rho_L u_L+p_L-p_R}{(S_R-u_R)\rho_R+(u_L-S_L)\rho_L}
\end{equation}
and the intermediate pressure
\begin{equation}
p^* = p_L+\rho_L(S_L-u_L)(S_M-u_L) = p_R+\rho_R(S_R-u_R)(S_M-u_R)
\end{equation}
which describe the interface condition. The minimum and maximum wave speed $S_L$ and $S_R$ are estimated by
\begin{equation}
S_L = \min[u_L-c_{s,L}, \tilde{u}-\tilde{c_s}], \quad S_R = \max[\tilde{u}+\tilde{c_s}, u_R+c_{s,R}].
\end{equation}
For two adjacent states described by different EOSs, the averaged speed of sound $\tilde{c_s}$ is obtained by
\begin{equation}
\tilde{c_s}^2 = \tilde{\Psi} + \tilde{\Upsilon} \widetilde{\left(\frac{p}{\rho}\right)},
\end{equation}
where the tilde on the right side indicates Roe-averaged values
\begin{equation}
\tilde{\rho} = \sqrt{\rho_L \rho_R}, \quad \tilde{f} = \mu(f)=\frac{\sqrt{\rho_L}f_L+\sqrt{\rho_R}f_R}{\sqrt{\rho_L}+\sqrt{\rho_R}}, f=\Psi, \Upsilon
\end{equation}
and
\begin{equation}
\widetilde{\left(\frac{p}{\rho}\right)} = \mu(\frac{p}{\rho})+\frac{1}{2} \tilde{\rho} \left(\frac{u_R-u_L}{\sqrt{\rho_L}+\sqrt{\rho_R}}\right)^2.
\end{equation}
Then we can calculate the interface flux between materials $\xi$ and $\eta$ according to the interface condition $(\mathbf{u}^*_{\xi \eta}, p^*_{\xi \eta})$ obtained above, where $\mathbf{u}^*_{\xi \eta} = u^*_{\xi \eta} \mathbf{n}_{\xi \eta}$. For a two-material-cell, the flux $\mathbf{\hat{X}}$ in Eq. (\ref{discretized_equ}) has the same form as that in Ref. \cite{hu2004interface}. The multi-material-cell, such as a three-material-cell in Fig. \ref{stencil}, has a more complex interface flux
\begin{eqnarray}
\mathbf{\hat{X}}(\Delta \Omega^{a}_{i,j}) &=& \mathbf{\hat{X}}(\Delta \Gamma_{ab}) + \mathbf{\hat{X}}(\Delta \Gamma_{ac}) = \mathbf{\hat{X}}(\Delta \Gamma_{\overline{OA}})+\mathbf{\hat{X}}(\Delta \Gamma_{\overline{OC}}) \nonumber \\
\mathbf{\hat{X}}(\Delta \Omega^{b}_{i,j}) &=& -\mathbf{\hat{X}}(\Delta \Gamma_{ab}) - \mathbf{\hat{X}}(\Delta \Gamma_{bc}) = -\mathbf{\hat{X}}(\Delta \Gamma_{\overline{OA}})-\mathbf{\hat{X}}(\Delta \Gamma_{\overline{OB}})
\nonumber \\
\mathbf{\hat{X}}(\Delta \Omega^{c}_{i,j}) &=& \mathbf{\hat{X}}(\Delta \Gamma_{bc}) - \mathbf{\hat{X}}(\Delta \Gamma_{ac}) = \mathbf{\hat{X}}(\Delta \Gamma_{\overline{OB}})-\mathbf{\hat{X}}(\Delta \Gamma_{\overline{OC}})
\end{eqnarray}
where the interface network inside the cell, $\Delta \Gamma_{\overline{OC}}$, are the line segments plotted in Fig. \ref{discretized_equ}(b). The interface condition is used to calculate the flux, such as
\begin{equation}
\mathbf{\hat{X}}(\Delta \Gamma_{\overline{OA}}) = \left[0, \ p^*_{ab}\Delta \Gamma_{\overline{OA}} \mathbf{n}_{ab}^x, \ p^*_{ab}\Delta \Gamma_{\overline{OA}} \mathbf{n}_{ab}^y, \ p^*_{ab}\Delta \Gamma_{\overline{OA}} \mathbf{n}_{ab} \cdot \mathbf{u}^*_{ab} \right]^T .
\end{equation}
Although this ``full interaction model'' is accurate to calculate the interface flux, it is time consuming due to the fact that explicit extraction of interface-network is necessary to calculate the length and normal direction of $\Delta \Gamma_{\overline{OA}}$, $\Delta \Gamma_{\overline{OB}}$ and $\Delta \Gamma_{\overline{OC}}$. And it is difficult to extend to 3D since the number of interface segments for each material may be excessive. Consequently, we use a reduced interaction model for a multi-material-cell. The basic idea is that only consider the interaction between the heaviest two materials inside a multi-material-cell. Assuming the mass of each material has a relation, $\alpha^a \rho_a > \alpha^b \rho_b > \alpha^c \rho_c$ in Fig. \ref{discretized_equ}, the reduced model is given by
\begin{eqnarray}
\mathbf{\hat{X}}(\Delta \Omega^{a}_{i,j}) &=& \mathbf{\hat{X}}(\Delta \Gamma_{ab}) \approx \mathbf{\hat{X}}(\Delta \Omega^a_{\overline{AC}}) \nonumber \\
\mathbf{\hat{X}}(\Delta \Omega^{b}_{i,j}) &=& -\mathbf{\hat{X}}(\Delta \Gamma_{ab}) \approx -\mathbf{\hat{X}}(\Delta \Omega^a_{\overline{AC}}),
\end{eqnarray}
where $\Delta \Omega^a_{\overline{AC}}$ is the material boundary of material $a$ represented by its local level-set. The normal direction $\mathbf{n}_{\xi \eta}$ in the full interaction model can be approximated by the normal of a particular material boundary. As shown in Fig. \ref{subcell_volume}(a), the three normal directions of material $a$, $b$ and $c$ are defined at the cell center and can be calculated by its corresponding local level-set function
\begin{equation}
\mathbf{n}_{\chi_r}= \frac{\nabla \phi^{r}}{|\nabla \phi^{r}|}, \quad \chi_r = a, b, c
\end{equation}
Note that the normal direction points into the respective material. Then the local coordinate of the Riemann problem $\mathsf{R}(W_L, W_R)$ is
\begin{equation}
\mathbf{n}_{\xi \eta} = \begin{cases}
\mathbf{n}_{\xi} \quad {\rm if}~ \alpha^{\xi} \rho_{\xi}  \geq \alpha^{\eta} \rho_{\eta}  \cr
\mathbf{n}_{\eta} \quad \rm{otherwise}	
\end{cases}.
\end{equation}
Additionally, the interface flux is modified by
\begin{equation}
\mathbf{\hat{X}}(\Delta \Omega^a_{\overline{AC}}) = \left[0, \ p^*_{ab}\Delta \Omega^a_{\overline{AC}} \mathbf{n}_{a}^x, \ p^*_{ab}\Delta \Omega^a_{\overline{AC}} \mathbf{n}_{a}^y, \ p^*_{ab} \Delta \Omega^a_{\overline{AC}} \mathbf{n}_{a} \cdot \mathbf{u}^*_{ab} \right]^T,
\end{equation}
where $\mathbf{n}_{a}$ is used to replace $\mathbf{n}_{ac}$ as $\alpha^a \rho_a > \alpha^c \rho_c$. The averaged flow variable of the two parts separated by $\Delta \Omega^a_{\overline{OA}}$ are $W_a$ and $W_b$, respectively, indicating the flow state of material $c$ is approximated by the data of material $b$. So we reduce the multiple times interaction to one time interaction, irrespectively of the number of materials inside a cell. And the explicit interface extraction is no longer needed to calculate the normal direction. Thus the reduced model significantly improves the efficiency, especially in 3D. The overall conservation is also achieved by summing for all materials
\begin{equation}\label{sum_equ2}
\sum_{\chi \in \mathrm{X}} \sum_{i,j} \alpha_{i,j}^{\chi,n+1}\mathbf{U}_{i,j}^{n+1} = \sum_{\chi \in \mathrm{X}} \sum_{i,j} \alpha_{i,j}^{\chi,n}\mathbf{U}_{i,j}^{n} + \mathrm{boundary} \ \mathrm{terms}
\end{equation}
as the interface fluxes for the two interacted materials have opposite sign in all multi-material-cells, i.e., $\mathbf{\hat{X}}(\Delta \Omega^{\xi}_{i,j})+\mathbf{\hat{X}}(\Delta \Omega^{\eta}_{i,j})=0$.
Note that the small volume fraction of a particular material may lead to numerical instability if the time step of explicit time integration schemes is calculated according to the full grid size CFL condition. In order to maintain numerical stability without reducing the time step, we apply the mixing procedure \cite{hu2006conservative} after each Runge-Kutta sub-step. For each material, the conservative quantities of small volume fraction cell are mixed with those of the larger neighboring cells in a conservative way. The exchanges of the conservative quantities $\mathbf{M}$ are calculated according to the averaged values, see Ref. \cite{hu2006conservative}. Then the conservative quantities for each material in the near interface cells are updated by
\begin{equation}
\alpha_{i,j}^{\chi,n+1}\mathbf{U}_{i,j}^{n+1} = (\alpha_{i,j}^{\chi,n+1}\mathbf{U}_{i,j}^{n+1})^* + \sum \mathbf{M}^x + \sum \mathbf{M}^y,
\end{equation}
where the second and third terms on the right hand side represent the sum of all mixing exchanges on cell $C_{i,j}$ in the $x$ and $y$ directions, respectively.
\subsubsection{Interface scale separation model}\label{separation-model}
\begin{figure}[p]
\begin{center}
\includegraphics[width=1.0\textwidth]{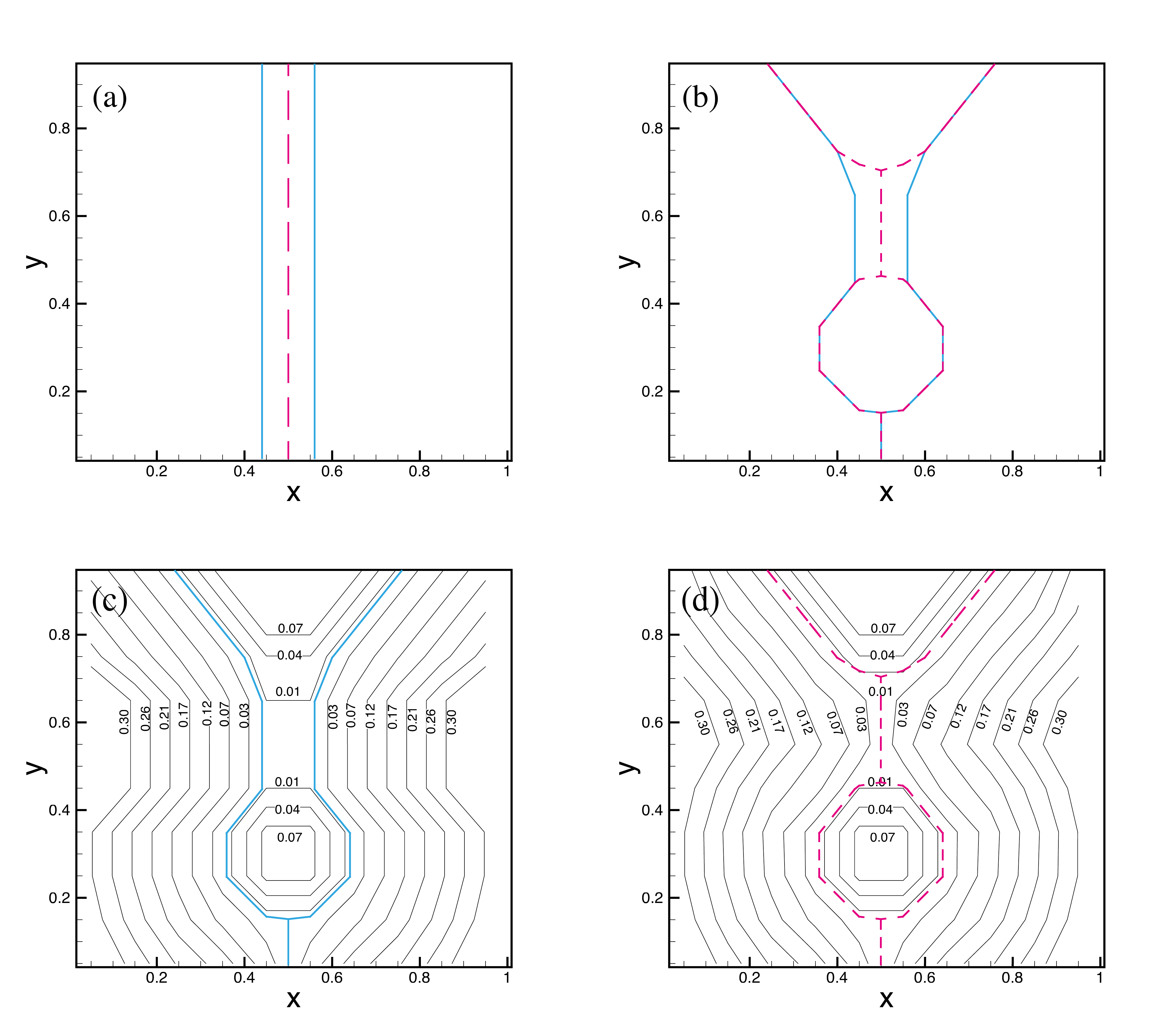}
\caption{Simple test cases for multi-material interface scale separation model: (a) A thin filament and (b) a small droplet. The solid blue line and dashed red line indicate the interfaces before and after applying the scale separation model. (c) and (d) show the level-set contours before and after applying the scale separation model.}
\label{multiscale}
\end{center}
\end{figure}

In this section, we discuss a numerical procedure for consistent removal of non-resolved interface segments during multi-material simulations. For a given spatial resolution non-resolved interfacial scales, such as thin filaments and small droplets, need to be removed in order to avoid proliferation of artifacts, so the separation of resolvable and non-resolvable interface scales are necessary. For non-resolved interface in two-material-cells, the scale separation operation is performed with the model in Refs. \cite{han2015scale, luo2016efficient} without modifications. While in multi-material-cells, non-resolved interface structures may occur across different material domains, so the scale separation model must have the adaptation to multi-material interfaces, i.e., after the removing, new interfaces need to be constructed between these materials to make sure those different materials keep separated with a different connection relationship. The basic idea of our interface scale separation model is same as the previous two models \cite{han2015scale, luo2016efficient}: although each material domain is simply connected for all resolved part, its sub-domain covered by the $\epsilon^-$-material-boundary which is obtained by slightly shifting the material boundary inward is not so for non-resolved part. This property is used to find "oddball cells" where non-resolved interface exists. We define the $\epsilon^\pm$-material-boundary as
\begin{equation}
\Gamma^\chi_{\epsilon^\pm} = \{\mathbf{x} \subset \Omega^\chi| \phi^\chi(\mathbf{x}) \pm \epsilon = 0, \chi \in \mathrm{X} \}
\end{equation}
where the $\epsilon$ is a positive constant \cite{han2015scale} and equals $0.75h$ for 2D and $0.9h$ for 3D to remove interface segment whose scales are smaller than grid scale $h$ \cite{han2015scale}. The set of cut cells $S_0$ which contain the segments of the interface network is defined as
\begin{equation}
S_0 = \{C_{i,j} | \partial C_{i,j} \cap \Gamma \neq\emptyset \}.
\end{equation}
For each material $\chi$ we define $S_{\pm}^{\chi}$ as the sets of cut cells containing the segments of $\partial \Omega^{\chi}_{\epsilon^{\pm}}$:
\begin{equation}
S_{\pm}^{\chi} = \{C_{i,j} | \partial C_{i,j} \cap \partial \Omega^\chi_{\epsilon^\pm} \neq\emptyset \}.
\end{equation}
Then the implementation of this model contains three main steps. First we identify the oddball cells where non-resolved interface segments reside. The selection criterion of oddball cells is similar with Ref. \cite{luo2016efficient}, which requires that each oddball cell can not find any neighbor that belongs to $S_{-}^{\chi_1}$:
\begin{equation}
S_{*}  = \{C_{i,j} | (C_{i,j} \in S_0) \wedge (\forall i_0, j_0 \in \{-1,0,1\}, C_{i+i_0,j+j_0} \not\in S_{-}^{\chi_1} ) \},
\end{equation}
where $\chi_1 = \chi_{i,j}$ is the primary indicator of $C_{i,j}$.
Second the local non-resolved topology of the oddball cell in $S_{*}$ is altered to a new resolved topology by replacing the its indicator with the indicator corresponding to second largest volume fraction
\begin{equation}
\chi_1^* = \underset{\chi_r \neq \chi_1}{\text{arg\, max}}\, \alpha_{\chi_r}
\end{equation}
as it has the largest possibility after scale separation. Note that this procedure automatically generates a new interface. Finally, the unsigned distance function of cell containing non-resolved interface segment is assigned with a reasonable new values with respect to the interface network of the new resolved topology. This is accomplished by the same operation in Ref. \cite{luo2016efficient} due to its efficiency. Because the considered cell $C_{i,j}$ in $S_{*}$ belongs to $\Omega^{\chi_1^*}$ after scale separation, we calculate the distance from its center to $\partial \Omega^{\chi_1^*}_{\epsilon^{+}}$
\begin{equation}
d = \sqrt{\left[i-i_0 \Delta x + (\varphi_{i_0,j_0} + \epsilon) n_x\right]^2+\left[j-j_0 \Delta y + (\varphi_{i_0,j_0} + \epsilon) n_y\right]^2},
\end{equation}
where $(i_0,j_0)$ is the index pair of the cell in $S_{+}^{\chi_1^*}$ within a search stencil $i-3\leq i_0 \leq i+3, j-3\leq j_0 \leq j+3$. $(n_x,n_y)$ are the unit normal vector at the center of $C_{i_0, j_0}$. Then we update the regional level-set with
\begin{equation}
\phi^*_{i,j} = (\varphi^*_{i,j}, \chi_1^*), \quad \varphi^*_{i,j} = |d_{min}-\epsilon|.
\end{equation}
Fig. \ref{multiscale} shows two simple test cases (a thin filament and a small droplet) where the non-resolved interface scales exist. The computational domain is a unit square with the grid size $\Delta x = \Delta y =0.1$. The thickness of the thin filament is $1.2 \Delta x$ and the radius of the small droplet is $1.5 \Delta x$.  In Fig. \ref{multiscale}(a), two interfaces that separates three materials merge to one after using the scale separation model, indicating that the material in the filament is entirely removed. The small droplet in Fig. \ref{multiscale}(b) becomes an isolated bubble and the filament connecting the droplet and the main body shrinks to a single interface while the other interface segments keep invariant. Meanwhile two additional triple points generate. The level-set contours keep regular after the scale separation and the contours inside the bubble are not changed by the separation operation, as shown in Figs. \ref{multiscale}(c) and \ref{multiscale}(d).
\subsection{Space-time adaptivity\label{adaption}}
To archive high computational efficiency and low memory storage, the space-time adaptivity strategy developed in Ref. \cite{han2014adaptive} is incorporated into our compressible multi-material method with minor changes. In detail, the multi-resolution method \cite{harten1995multiresolution} is used for mesh refinement due to its high rate of data compression. The projection and prediction operators \cite{roussel2003conservative} are defined based on the cell-averaged multi-resolution representation. For simplicity, the 1D operators with 5th-order interpolation are
\begin{equation}
P_{\ell+1 \rightarrow \ell}: \quad \bar{u}_{\ell,i} = \frac{1}{2} (\bar{u}_{\ell+1,2i}+\bar{u}_{\ell+1,2i+1}),
\end{equation}
and
\begin{eqnarray}
P_{\ell \rightarrow \ell+1}: \quad \hat{u}_{\ell+1,2i} &=& \bar{u}_{\ell,i} + \sum_{m=1}^{2}\gamma_m (\bar{u}_{\ell,i+m}+\bar{u}_{\ell,i-m}),
\\ \nonumber
\hat{u}_{\ell+1,2i+1} &=& \bar{u}_{\ell,i} - \sum_{m=1}^{2}\gamma_m (\bar{u}_{\ell,i+m}+\bar{u}_{\ell,i-m}),
\end{eqnarray}
where $\ell$ is the index of levels and $\gamma_m$ is the interpolation coefficient. The the mesh refinement and coarsening are accomplished by comparing the prediction error $\bar{d}_{\ell,i} = \bar{u}_{\ell,i} - \hat{u}_{\ell,i}$ with a level-dependent threshold \cite{han2014adaptive}. A multi-step Runge-Kutta local time stepping scheme \cite{domingues2008adaptive} is employed to archive time adaptivity and thus obtain additional speed-up. To maintain strict conservation a conservative flux correction \cite{domingues2008adaptive} is adopted between cells with different levels.

\begin{figure}[p]
\begin{center}
\includegraphics[width=0.95\textwidth]{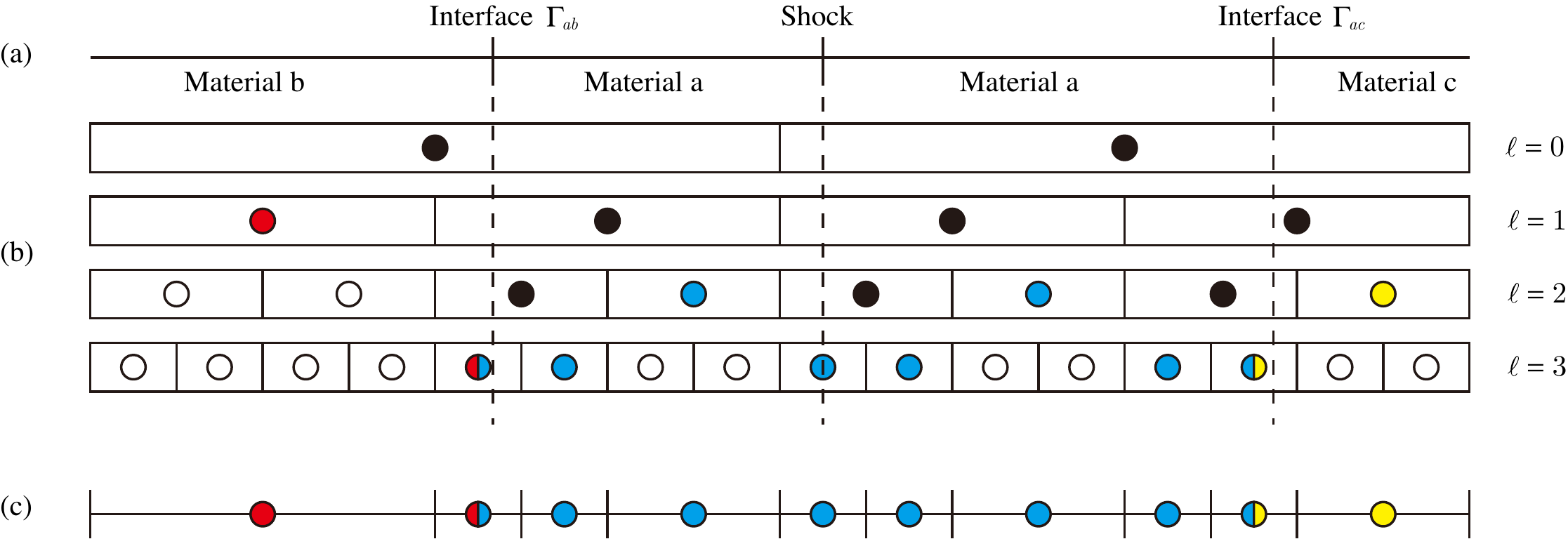}
\caption{The multi-resolution representation of 3-material compressible flows in 1D.
(a) The computational domain is partitioned by 3 materials ($a$, $b$, $c$). Two interfaces $\Gamma_{ab}$ and $\Gamma_{ac}$ separate these materials, and a shock wave occurs inside material $a$.
(b) The circles indicate the non-existing blocks. The black dots are the non-leaf blocks which have child blocks. The leaf blocks, colored by blue, red and yellow are blocks containing single material $a$, $b$ and $c$, respectively.
(c) The final multi-resolution representation shows the single-material block of material $b$ is located at the coarsest level as no shock wave or interface occur near it. The single-material block of material $c$ is refined to $\ell=2$ because it is close to the interface $\Gamma_{ac}$. The single-material blocks of material $a$ are either refined by interface or shock wave. All multi-material blocks are refined to the finest level.}
\label{multiresolu}
\end{center}
\end{figure}

The pyramid data structure and storage-and-operation-splitting approach proposed in Ref. \cite{han2014adaptive} are used here. The block containing cells which reside in the narrow band of the interface network is refined to the finest level $\ell_{max}$ and denoted to ``multi-material block'', otherwise to ``single-material block''. In such way the operation related to interface, including interface interaction, mixing procedure, scale separation and level-set advection, are only conducted at the finest level. The block position identifier \cite{han2014adaptive} is used to distinguish the block location. The identifier at the finest level is 1 whenever the block has cells which occur in the narrow band of any cut cell, otherwise it is 0. A cell $C_{i,j}$ contains interface network if it is intersected by the zero contour of local level-set field
\begin{equation}
\exists r \in \mathrm{X}_s, C_{i,j} \cap \{\mathbf{x} | \phi^r(\mathbf{x}) =0 \} \neq \emptyset,
\end{equation}
or the indicator field in $V_s$ is different with $\chi_{i,j}$,
\begin{equation}
\exists C_{k,l} \in V_s, \chi_{k,l} \neq \chi_{i,j}.
\end{equation}
Then the position identifier of blocks at other levels are obtained according to Ref. \cite{han2014adaptive}. The final multi-resolution representation of a multi-material problem is generated by locations of interface and shock waves, see Fig. \ref{multiresolu} for example.

Note that the ``single-material block'' may also exist at the finest level if shock wave resides in this block. For a ``multi-material block'', we allocate memory for one single regional level-set field and $\mathcal{N}_b$ flow state fields, where $\mathcal{N}_b$ is determined by searching all unique indicator inside the inner and buffer zone of this block and has the relation $\mathcal{N}_b \ll \mathcal{N}$ when $\mathcal{N}$ is large. For a ``single-material block'', the volume fraction and apertures become unit and the interface exchange terms vanish. Thus the governing equation Eq. (\ref{discretized_equ}) degenerates to standard finite volume scheme on a 2D Cartesian grid.
\section{Numerical validation \label{sec:validation}}
In this section, we assess the accuracy and robustness of present method by a number of 1D and 2D test cases. First, the 1D multi-material shock tube and ICF implosion are tested. Then more complex 2D cases, including 2D ICF implosion, compressible triple point and shock wave interactions in multiple materials
serve to demonstrate the robustness of interface interaction model and multi-material scale separation method in high-resolution simulations. For all test cases, the fluid dynamics and interface advection are solved by a 5th-order WENO \cite{jiang1995efficient} and a 2nd-order TVD Runge-Kutta scheme \cite{shu1988efficient}, with a CFL number of 0.6.

%
\begin{table}
\centering
\resizebox{\textwidth}{!}{
\begin{threeparttable}
\caption{\label{table1} Initial conditions for 1D test cases.}
\vspace{0.5cm}
\begin{tabular}{|l|ccc|ccc|}
\hline
\multirow{3}{*}{} & \multicolumn{3}{c|}{Case I\tnote{*}}  & \multicolumn{3}{c|}{Case II}\\
\hline
Location &$0\leq x<0.25$ & $0.25\leq x<0.5$ & $0.5\leq x \leq 1.0$ & $0\leq r<1.0$ & $1.0\leq r<1.2$ & $1.2\leq r \leq 1.5$\\
\hline $\chi$ & 1	& 2	& 3		& 1	& 2	& 3\\
 $\rho$  & 0.125	& 1.0 & 0.125		&0.05	&1.0	&0.1\\
 $p$  & 0.1	& 1.0 & 0.1		&0.1	&0.1	&13.0\\
 $\gamma$  & 1.667	& 1.4 & 1.667	&1.667	&1.667	&1.667\\
\hline
\multicolumn{7}{l}{%
  \begin{minipage}{16.5cm}%
    \small Note: the velocity $u$ is zero everywhere.%
  \end{minipage}%
}\\
\end{tabular}
    \begin{tablenotes}
    \small \item[*]{For the helium-air-R22 shock tube problem, just change $\gamma$ to $1.249$ in $0.5\leq x \leq 1.0$.}
    \end{tablenotes}
\end{threeparttable}
}
\end{table}
\subsection{Shock-tube problem (I) \label{shock-tube}}
\begin{figure}[p]
\begin{center}
\includegraphics[width=1.0\textwidth]{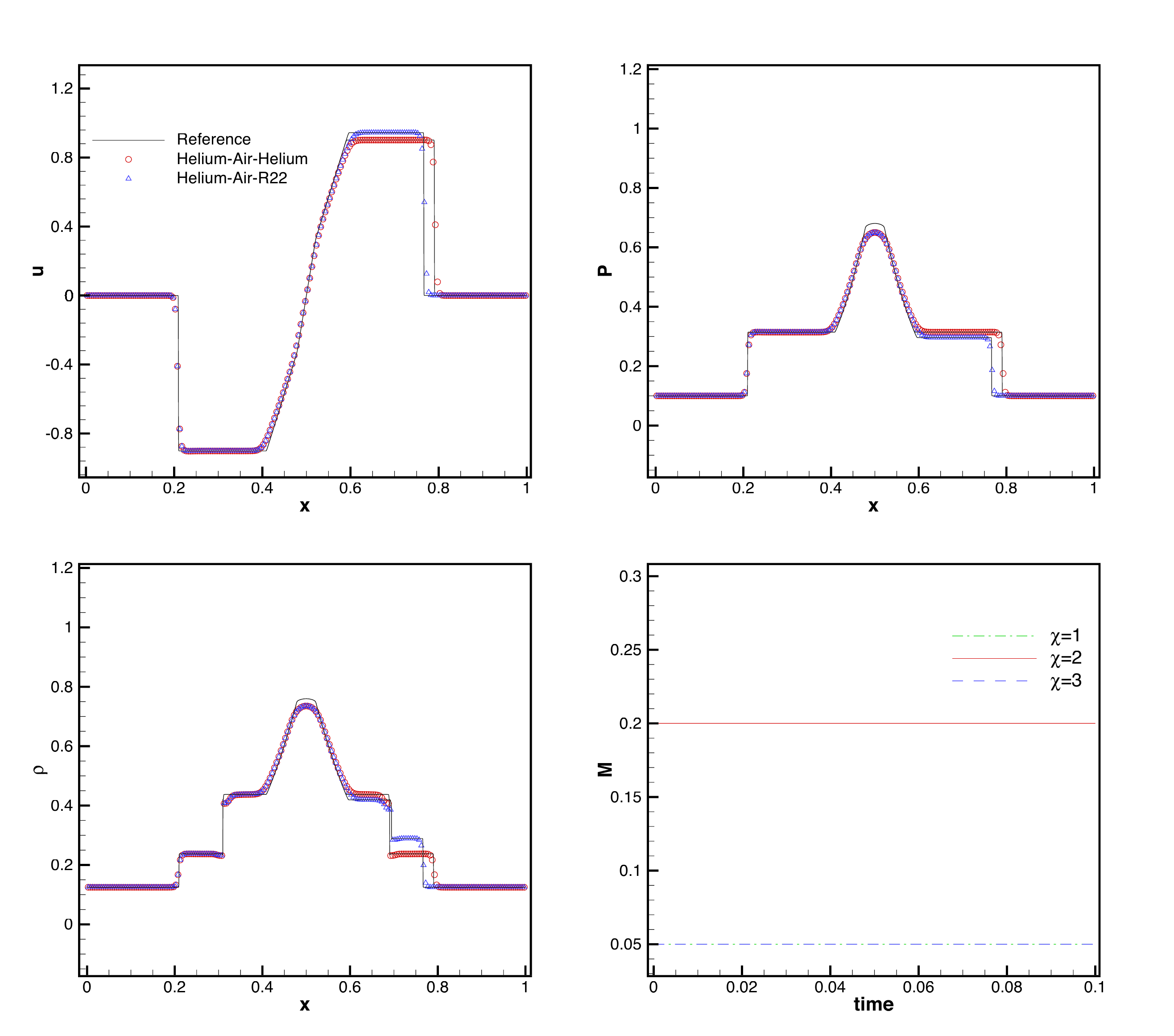}
\caption{There-material shock tube problem.}
\label{helium-air-helium}
\end{center}
\end{figure}

Three-material shock-tube problem of two helium gases and one air gas modeled by ideal-gas EOSs are simulated by our method. This case is a extension of the two-material shock-tube problem in Refs. \cite{hu2006conservative, fedkiw1999non}. Reflective boundary conditions are applied at $x = 0$ and $x = 1$. The initial condition is listed in Table. \ref{table1}. The grid spacing is $\Delta x = 5.0 \times 10 ^{-3}$ and the reference solution is a high-resolution numerical result with $\Delta x = 2.5 \times 10 ^{-4}$. Initially, with this setup, two Riemann problems occur at $x=0.4$ and $x=0.6$ and generate symmetric wave types. Two shock waves moves towards the left and right boundaries while two rarefaction waves are approaching to each other. Finally at $t=0.1$ these two rarefaction waves impact and interact with each other, as shown in Fig. \ref{helium-air-helium}. Good agreement with the reference solution is observed and the distributions of flow variables exhibit symmetric profiles. Then we modify the ratio of specific heats in $0.5\leq x \leq 1.0$ to R22's value, $\gamma = 1.249$. This treatment leads to a slightly asymmetry in the profiles, as shown in Fig. \ref{helium-air-helium}. More importantly, the mass of each individual material is strictly conserved during the simulation.
\subsection{1D Cylindrical ICF implosion (II) \label{1D_ICF}}
The setup of our implosion simulations in cylindrical geometry is taken from Ref. \cite{galera2010two}. The computational domain is $[0,1]$ and 1D axisymmetrical grids with different resolutions are used. There are three materials in the domain. A light fluid is located in core region of the target and is surrounded by a shell of dense fluid. Outside the shell is an ambient material which is not solved during the simulation. Thus we treat the interface between the shell and the ambient material as a free surface boundary. To drive the implosion the pressures $p(t)$ imposed on that boundary are initially constant and then decrease linearly as
\begin{equation}
p(t) = 13-\frac{12.5 (t-0.04)}{0.125-0.04},
\end{equation}
according to Ref. \cite{galera2010two}.

\begin{figure}[p]
\begin{center}
\subfloat[][]{\label{1D_ICF_R:a}\includegraphics[scale=0.25]{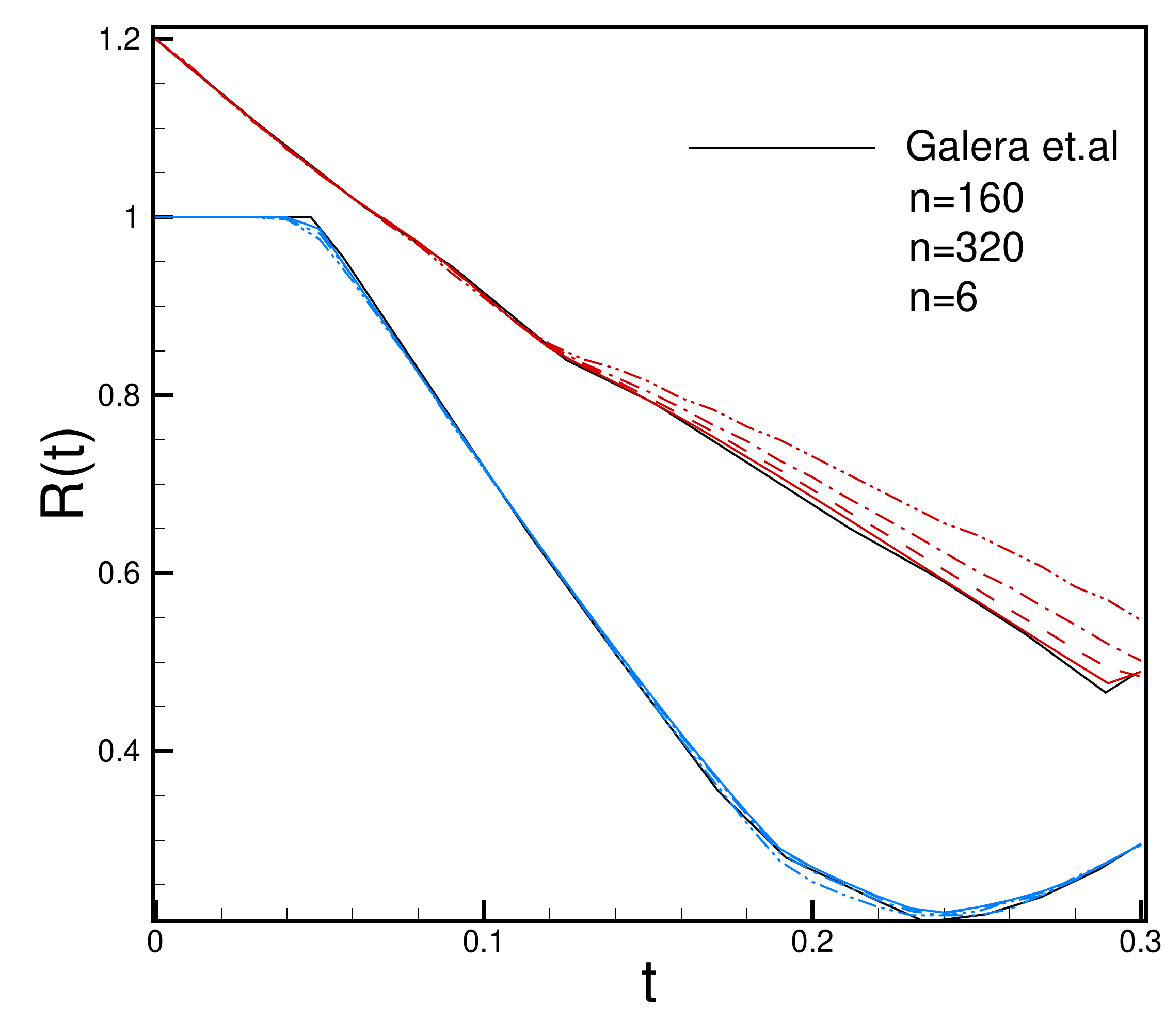}}
\caption{Trajectories of the inner (blue lines) and outer (red lines) interfaces of the shell in 1D ICF implosion problems.}
\label{1D_ICF_R}
\end{center}
\end{figure}

\begin{figure}[p]
\begin{center}
\subfloat[][]{\label{1D_ICF_R_para:a}\includegraphics[scale=0.25]{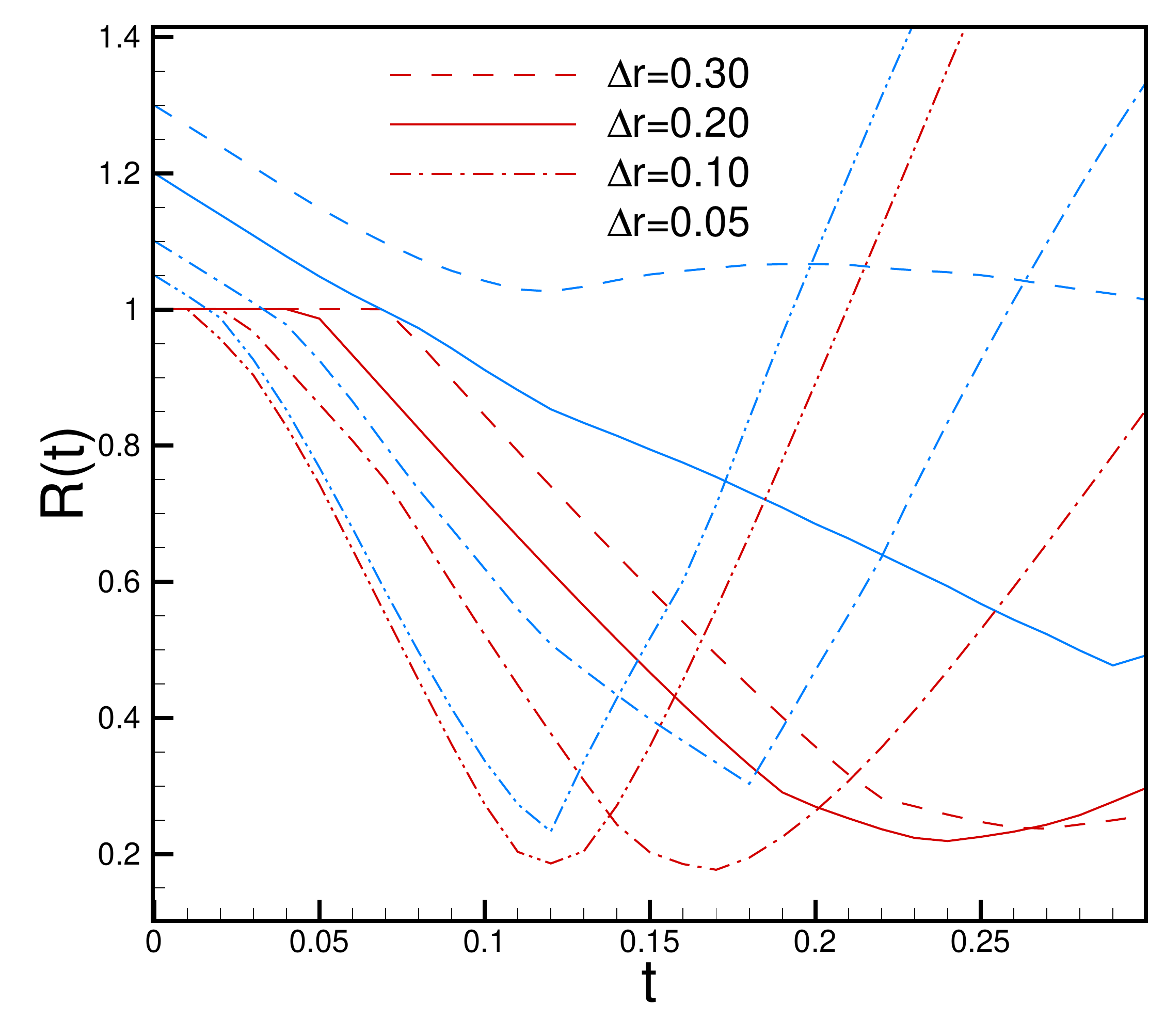}}
\subfloat[][]{\label{1D_ICF_R_para:b}\includegraphics[scale=0.25]{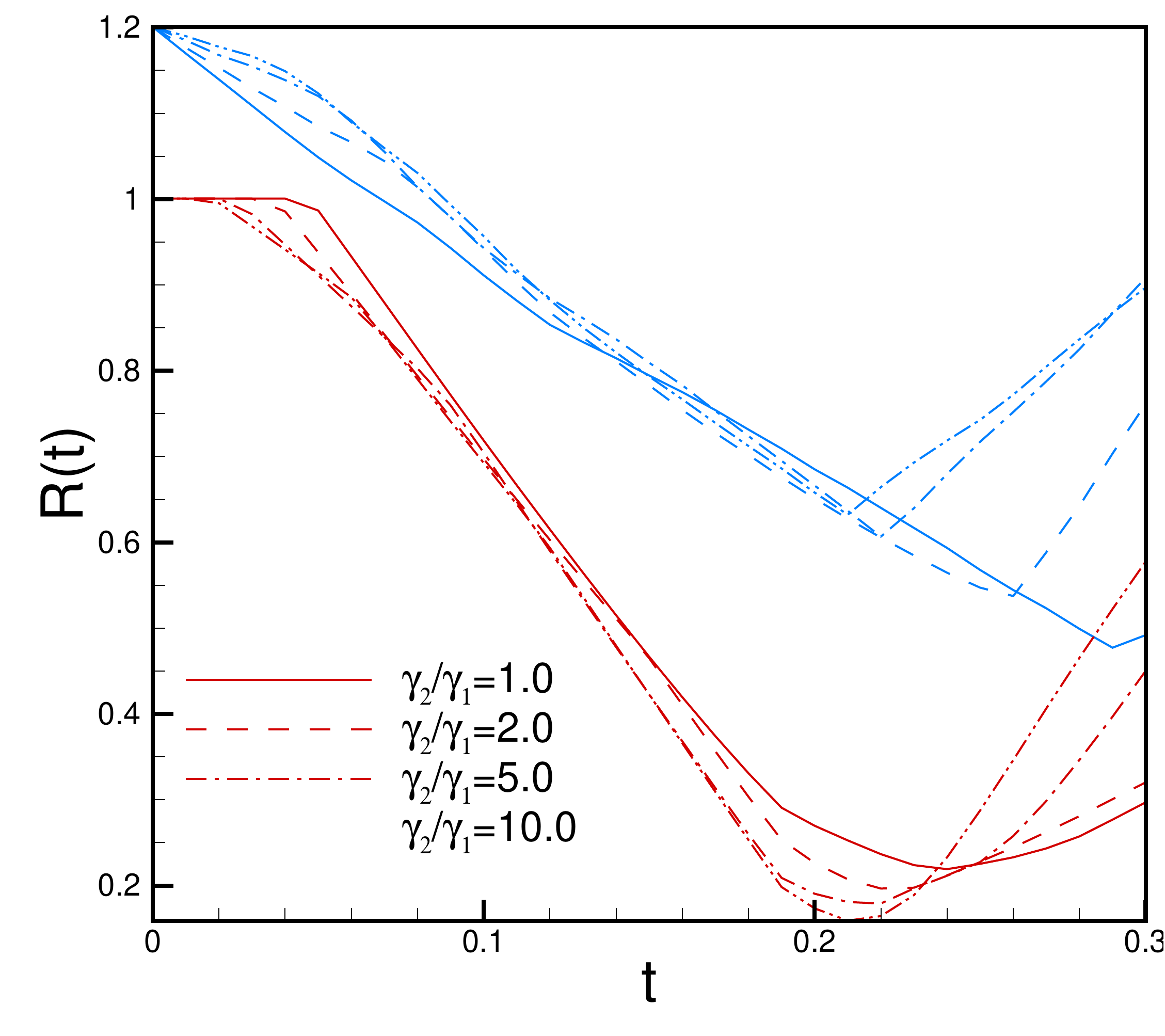}}
\caption{Parameter study of the ICF implosion problem: (a) different shell thickness and (b) different $\gamma_2 / \gamma_1$.}
\label{1D_ICF_R_para}
\end{center}
\end{figure}

The locations of inner and outer interfaces of the shell are plotted in Fig. \ref{1D_ICF_R}. In Fig. \ref{1D_ICF_R}(a), our results converge to the Lagrangian result \cite{galera2010two} with the cell number increasing from $160$ to $1280$. In the initial stage the outer interface moves inward under a constant driven pressure, leading to a shrinking of the shell until $t=0.047$. Then both the interfaces move towards the core. As the contraction speed of the outer interface is larger than that of the inner one, the thickness of the shell increases. The deceleration of the light fluid is observed from $t=0.17$ to $t=0.24$ when the radius of inner interface reaches its minimum value (referred to as ``stagnation time'' \cite{galera2010two}). Afterwards the light fluid exhibits a expansion. In order to demonstrate the high flexibility of our method, we conduct multiple simulations with varying parameters, $r_2$ and $\gamma_2$, which are the radius of the outer interface and the ratio of specific heats of the heavy fluid, respectively. This simple parameter study is useful in designing a ICF capsule as the thickness of shell and EOS used in the previous simulations \cite{galera2010two} are not realistic. As shown in Fig. \ref{1D_ICF_R_para}(a), the thickness of the shell during the simulation, the stagnation time and the compression rate reduce with the initial thickness $\Delta r = r_2 - r_1$ decreasing from $0.30$ to $0.05$. When the ratio $\gamma_2/\gamma_1$ varies from $1.0$ to $10.0$, the profiles for the inner and outer interfaces are shifted downwards and upwards, respectively, as plotted in Fig. \ref{1D_ICF_R_para}(b). Thus the archived minimum radius of inner interface decrease as the $\gamma_2/\gamma_1$ increase, indicating that stiff shell materials generate high compression rate.
\subsection{Compressible triple point problem (III) \label{2D_triple}}
In this section we simulate the compressible triple point problem which contains three perfect gases and is usually used to validate the accuracy and robustness of Lagrangian or ALE methods in simulations of mutli-material compressible flows \cite{galera2010two, zeng2014frame}. The computational domain is a rectangle $\left[0, 7\right]\times\left[0, 3\right]$ and is partitioned into three sub-domains: (i) $\left[0, 1\right]\times\left[0, 3\right]$ which is filled with a high pressure high density fluid $(\rho, p, \gamma, \chi) = (1, 1, 1.5, 1)$, (ii) $\left[1, 1\right]\times\left[0, 1.5\right]$ which is occupied by a low pressure high density fluid $(\rho, p, \gamma, \chi) = (1, 0.1, 1.4, 2)$, and (iii) $\left[1, 7\right]\times\left[1.5, 3\right]$ which has a low pressure low density flow state $(\rho, p, \gamma, \chi) = (0.125, 0.1, 1.5, 3)$. Accordingly, a triple point exists initially at $(1,1.5)$. Reflective boundary conditions are employed and the final time of the simulation is $5.0$. The coarsest level has $7 \times 3$ blocks and are refined to the finest level by the mesh refinement criterion in Sec. \ref{adaption}.

\begin{figure}[p]
\begin{center}
\includegraphics[width=1.0\textwidth]{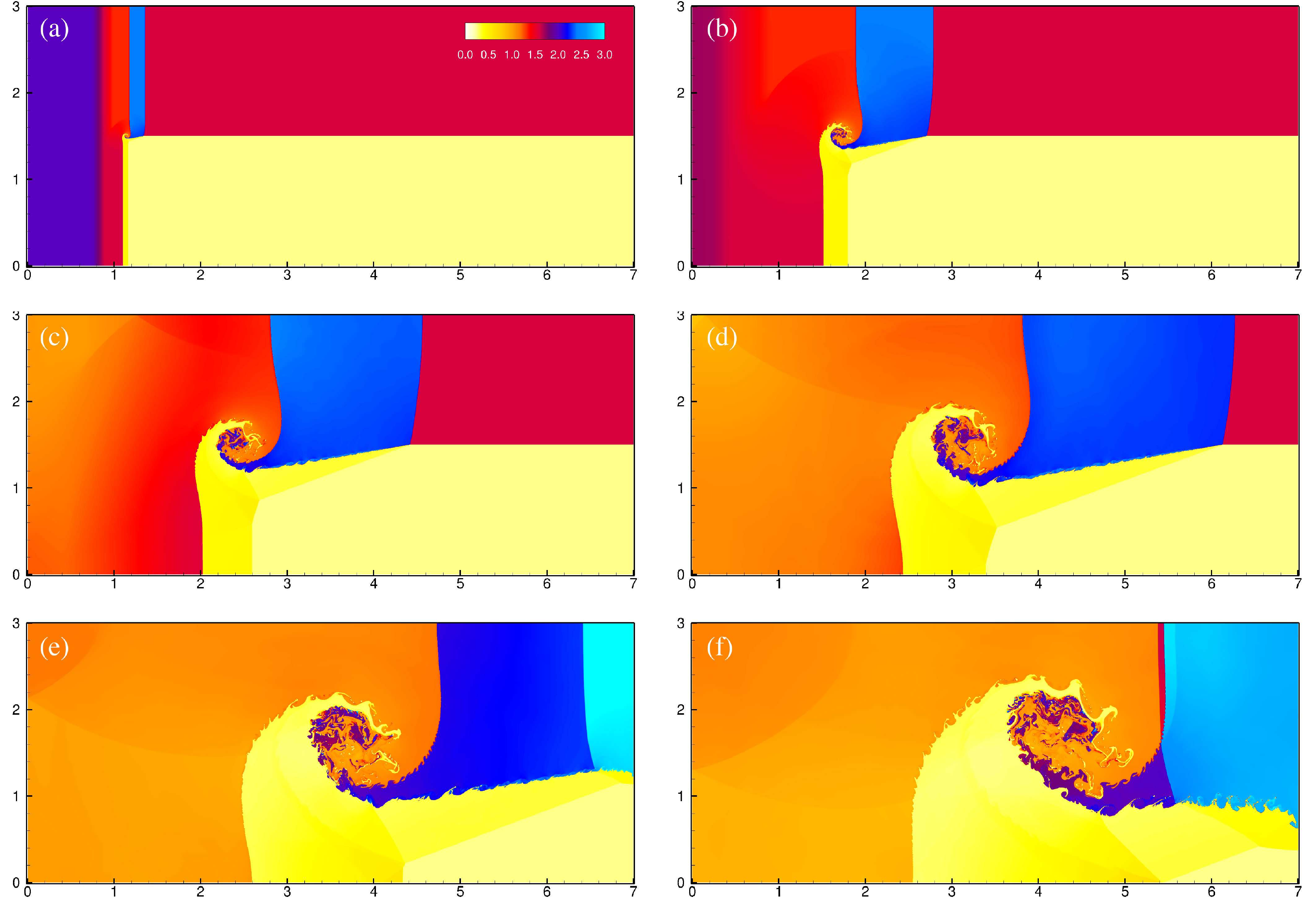}
\caption{Internal energy distributions at $t=0.2$, $1.0$, $2.0$, $3.0$, $4.0$ and $5.0$ for 2D compressible triple point problem with $\ell = 5$.}
\label{shock-triple-e}
\end{center}
\end{figure}

\begin{figure}[p]
\begin{center}
\includegraphics[width=1.0\textwidth]{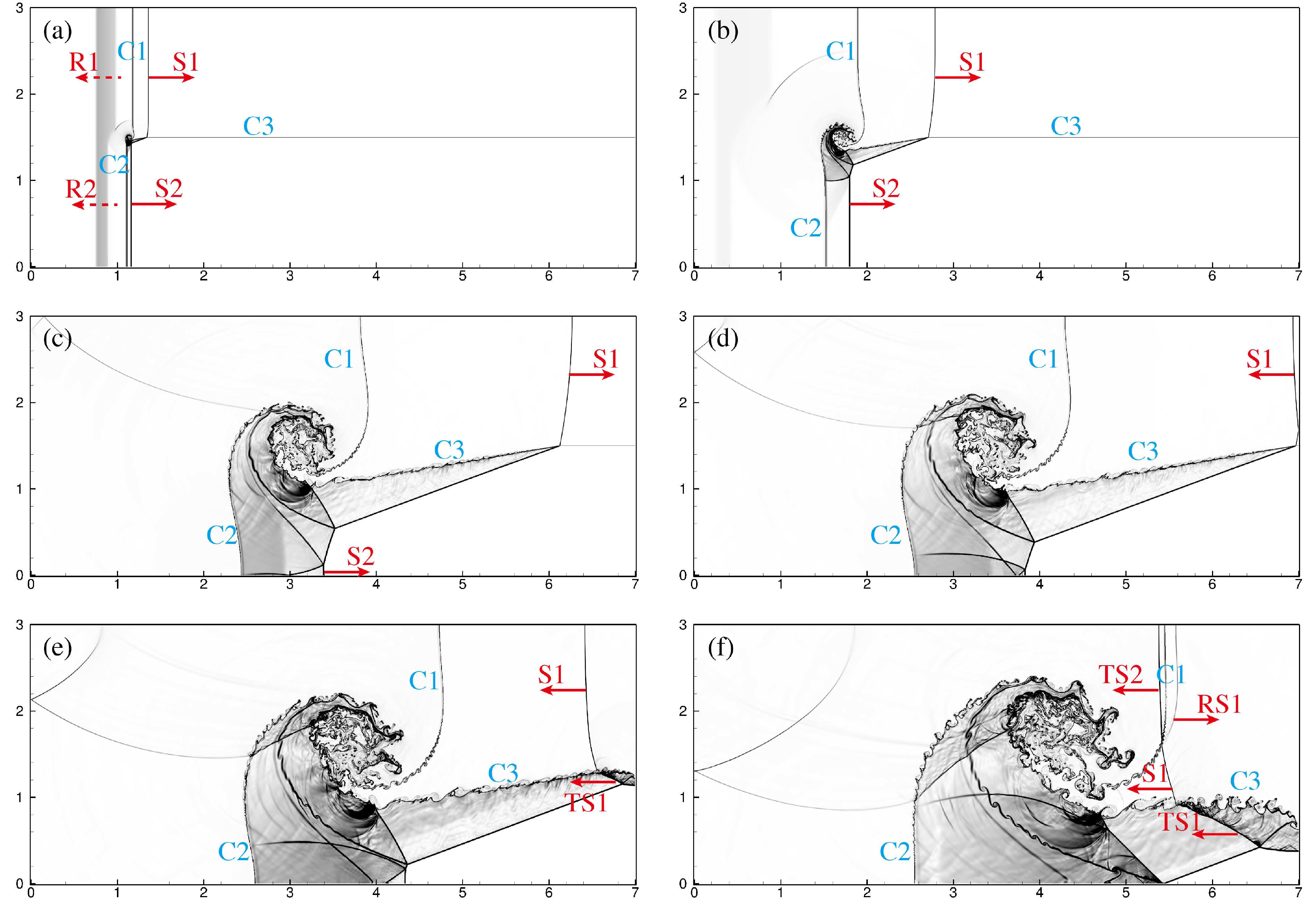}
\caption{Snapshots of density gradient at $t=0.2$, $1.0$, $3.0$, $3.5$, $4.0$ and $5.0$ for 2D compressible triple point problem with $\ell = 5$.}
\label{shock-triple-dr}
\end{center}
\end{figure}

\begin{figure}[p]
\begin{center}
\includegraphics[width=1.0\textwidth]{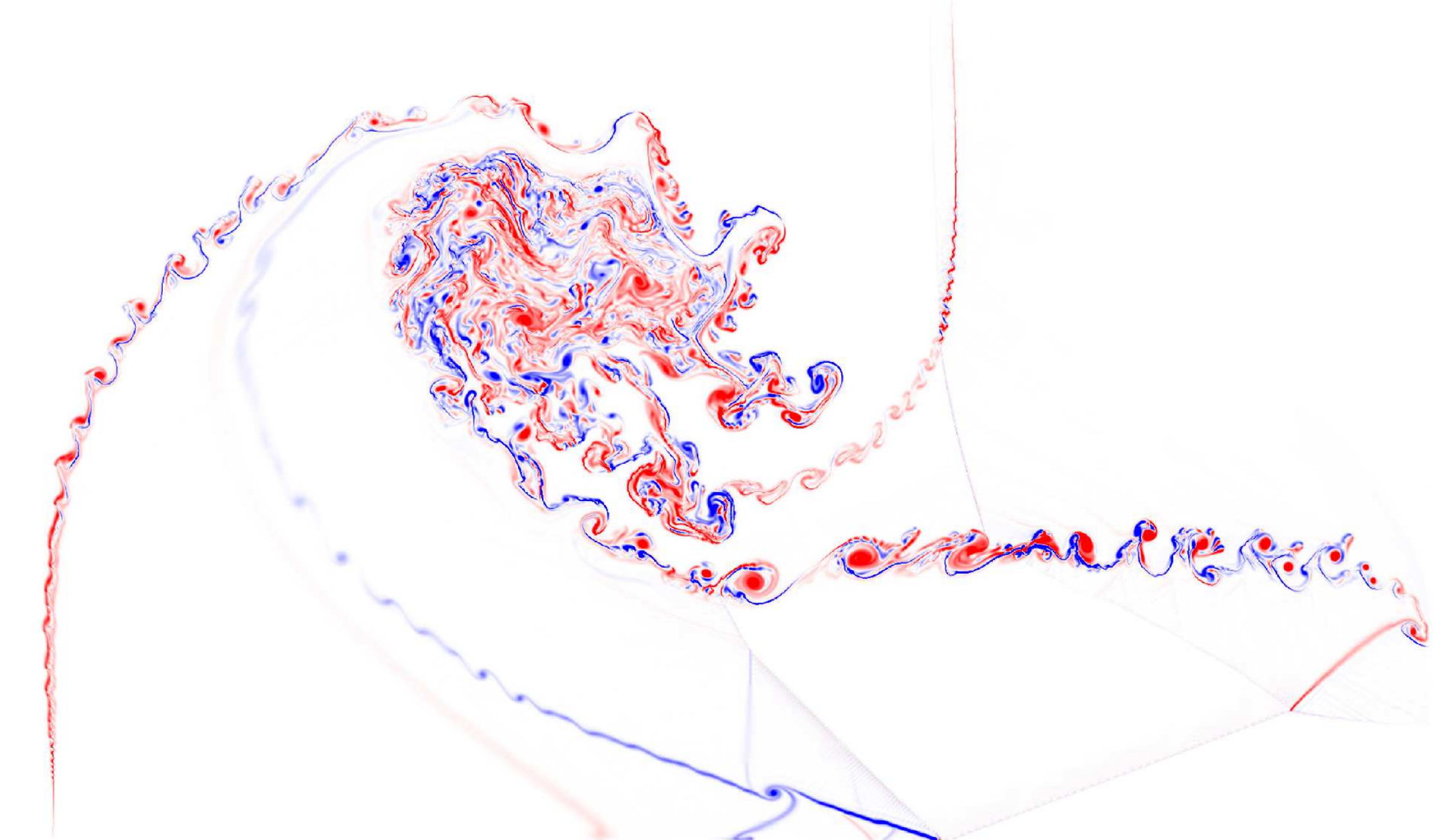}
\caption{Vorticity contours at $5.0$ for 2D compressible triple point problem with $\ell = 5$.}
\label{shock-triple-vor}
\end{center}
\end{figure}

First, we show the numerical results of a high-resolution simulation conducted with $\ell_{max} = 7$ and an effective resolution of $3584 \times 1536$ at the finest level. In Fig. \ref{shock-triple-e}, the internal energy contours at $t=0.2$, $1.0$, $3.0$, $3.5$, $4.0$ and $5.0$ are in good agreement with the numerical results in previous literatures \cite{galera2010two, kucharik2014conservative}. The development of the shock system is illustrated in Fig. \ref{shock-triple-dr}. Initially, both the solutions of the Riemann probelms at discontinuities $\Gamma_{13}$ and $\Gamma_{12}$ are composed of a contact discontinuity ($C1$ or $C2$), a leftward rarefaction wave ($R1$ or $R2$) and a rightward shock wave ($S1$ or $S2$), see Fig. \ref{shock-triple-dr}(a). The shock $S1$ moves faster than $S2$ as the acoustic impedance has a relation $\rho_2 c_{s,2}>\rho_3 c_{s,3}$. As a consequence, a distinct roll-up region formulates around the triple point, as shown in Fig. \ref{shock-triple-dr}(b). Besides, near the triple point the shock reflection pattern is more complex as different waves interact with each other. Meanwhile, the perturbation development is observed in Fig. \ref{shock-triple-dr}(c) due to the strong shear along all the contact discontinuities $C1$, $C2$ and $C3$, which is not observed in numerical results of previous papers \cite{galera2010two, kucharik2014conservative} due to the high dissipation and low robustness in high-resolution simulations. Fig. \ref{shock-triple-dr}(d) shows the shock wave system just after $S1$ impacts the right boundary and reflects. At this time instant, the interface inside the roll-up is extensively perturbed, leading to numerous small droplets. Afterwards, $S1$ moves upstream and is partially refracted to generate a transmitted shock $TS1$, as shown in Fig. \ref{shock-triple-dr}(e). After $S1$ reaches contact discontinuity $C1$, it becomes a transmitted shock $TS2$ inside the material $1$ and a reflected shock $RS1$ is produced to maintain the mechanical equilibrium at the interface $\Gamma_{13}$, see Fig. \ref{shock-triple-dr}(f). Due to the Kelvin-Helmholtz instabilities along the contact discontinuities, a large number of vortical structures are produced during the simulations, as shown in Fig. \ref{shock-triple-vor}. Unlike the previous numerical results \cite{galera2010two, kucharik2014conservative}, the roll-up core contains the majority of the small scale features, indicating the filament inside the roll-up breaks up very quickly as our sharp-interface does not introduce large numerical dissipation near the interfaces.

\begin{figure}[p]
\begin{center}
\includegraphics[width=1.0\textwidth]{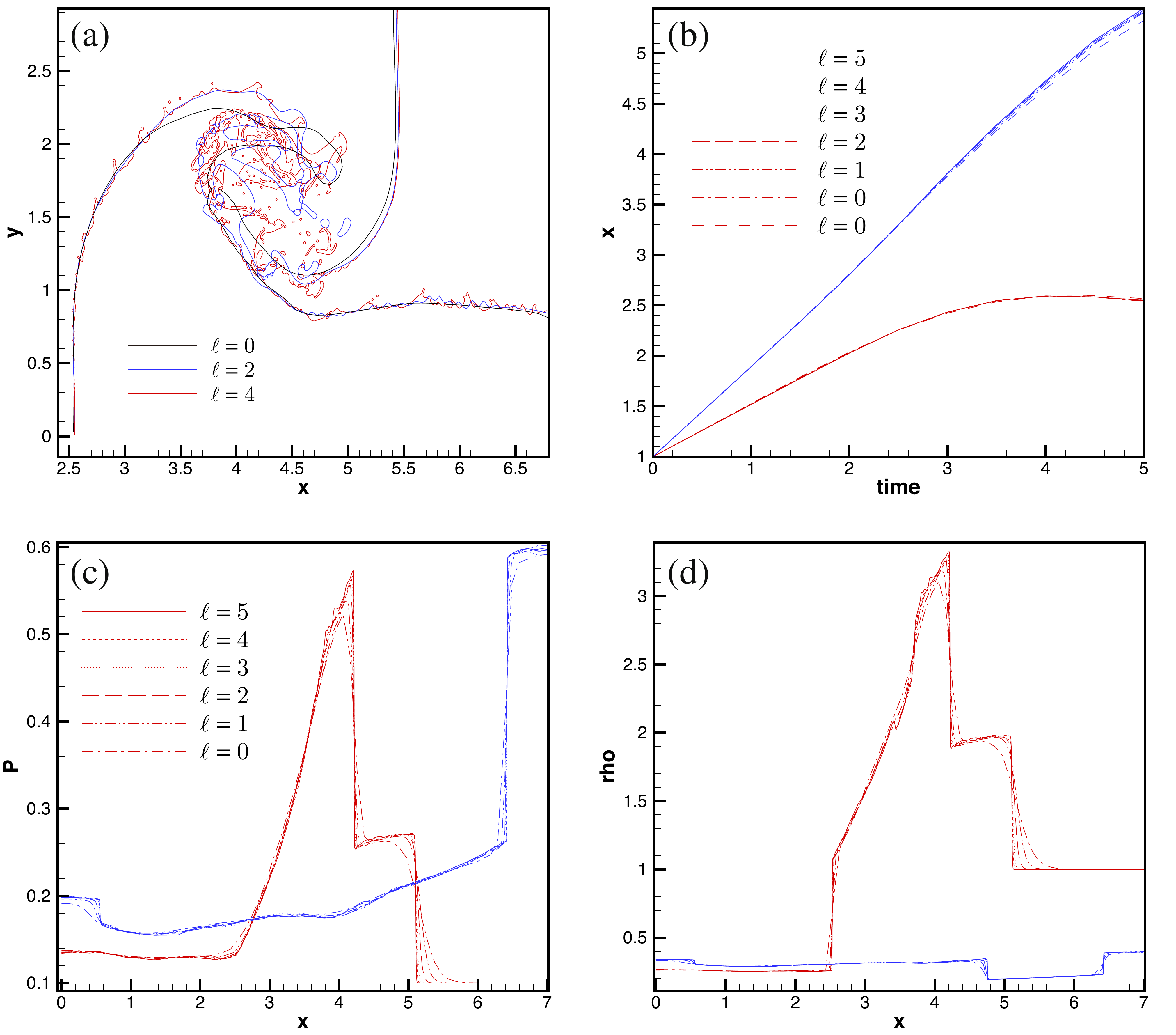}
\caption{Grid convergence tests for 2D compressible triple point problem. (a) The interface networks at $t=5.0$ with different finest levels, $\ell=0$, $\ell=2$ and $\ell=4$. (b) Convergence study of the interface location at the lower (red lines) and upper (blue lines) boundaries with increasing $\ell$ from $0$ (the second resolution with $\ell=0$ is half of the first one) to $5$. The coarsest resolution is obtained with $\ell=0$ and the number of inner cells is $8$ while in other simulations the number of inner cells is $16$. (c) The pressure profiles along $x$ direction at $y=0.5$ (red lines) and $y=2.5$ (blue lines) with $t$ being $4.0$.}
\label{shock-triple-profile}
\end{center}
\end{figure}

Furthermore, the grid convergence study in Fig. \ref{shock-triple-profile}(a) shows the vertical interfaces converges very quickly as the shear strength is small while horizontal interfaces are successively perturbed by the Kelvin-Helmholtz instabilities. As shown in Fig. \ref{shock-triple-profile}(b), the interface locations at the upper and lower boundaries exhibit a good grid convergence as the interfaces are strictly horizontal. At $t=4.0$, the shock waves and contact discontinuities corresponding to Fig. \ref{shock-triple-dr}(e) are represented by the pressure and density profiles along the $x$ direction at $y=0.5$ and $y=2.5$, see Figs. \ref{shock-triple-profile}(c) and (d). The pressure and density jumps are successively sharpened with increasing resolution from $\ell_{max} = 0$ to $\ell_{max} = 5$ and do not show any overshot in all profiles.

\subsection{2D cylindrical ICF implosion with perturbed interface (IV) \label{2D_ICF}}
Following the Sec. \ref{1D_ICF}, we investigate 2D cylindrical ICF implosion problems with initial perturbations. The interface between light and heavy fluid has a single-mode perturbation \cite{galera2010two, joggerst2014cross} with
\begin{equation}
r_0' = r_0 [1+A \cos(m \theta)],
\end{equation}
where $r_0$ and $r_0'$ are the initial unperturbed and perturbed radii of the light fluid, and $\theta$ is the polar coordinate. The mode number $m$ is $5$ and $47$ for low-mode and high-mode perturbations, respectively. The amplitude $A$ of the initial perturbation is $2\%$ of the wavelength of low-mode perturbation. Free surface boundary condition is prescribed at the outer interface. There is one block at the coarsest level and the maximum level is $\ell_{max} = 8$. With each block containing $16 \times 16$ inner cells, the effective grid resolution at the finest level is $4096^2$. Other computational setup is the 2D extension of the 1D case in Section \ref{1D_ICF}.

\begin{figure}[p]
\begin{center}
\includegraphics[height=12cm]{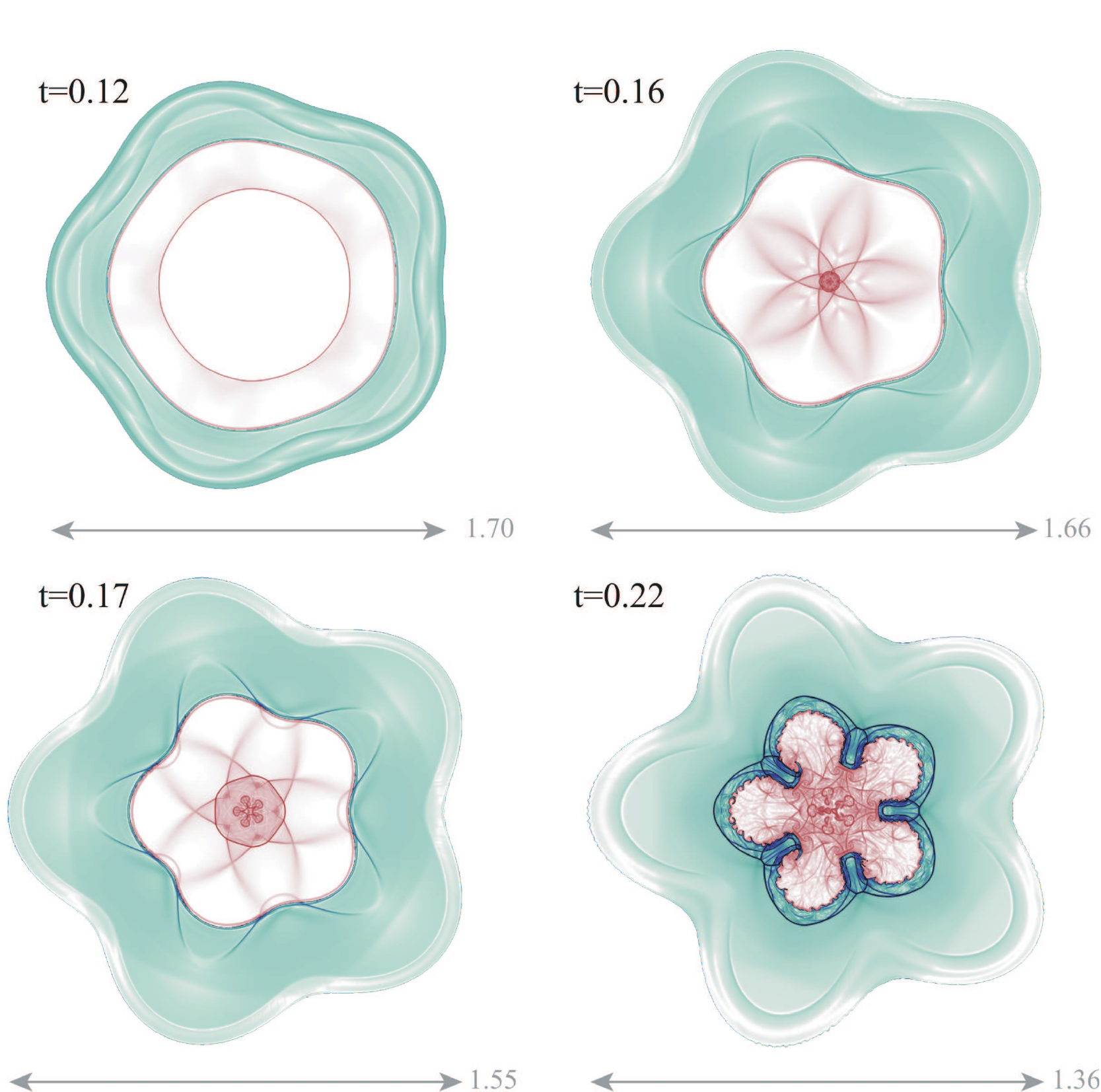}
\caption{Snapshots of density gradient before the stagnation time for 2D cylindrical ICF implosion with a 5-mode perturbation.}
\label{2D_ICF_low_1}
\end{center}
\end{figure}

\begin{figure}[p]
\begin{center}
\includegraphics[height=12cm]{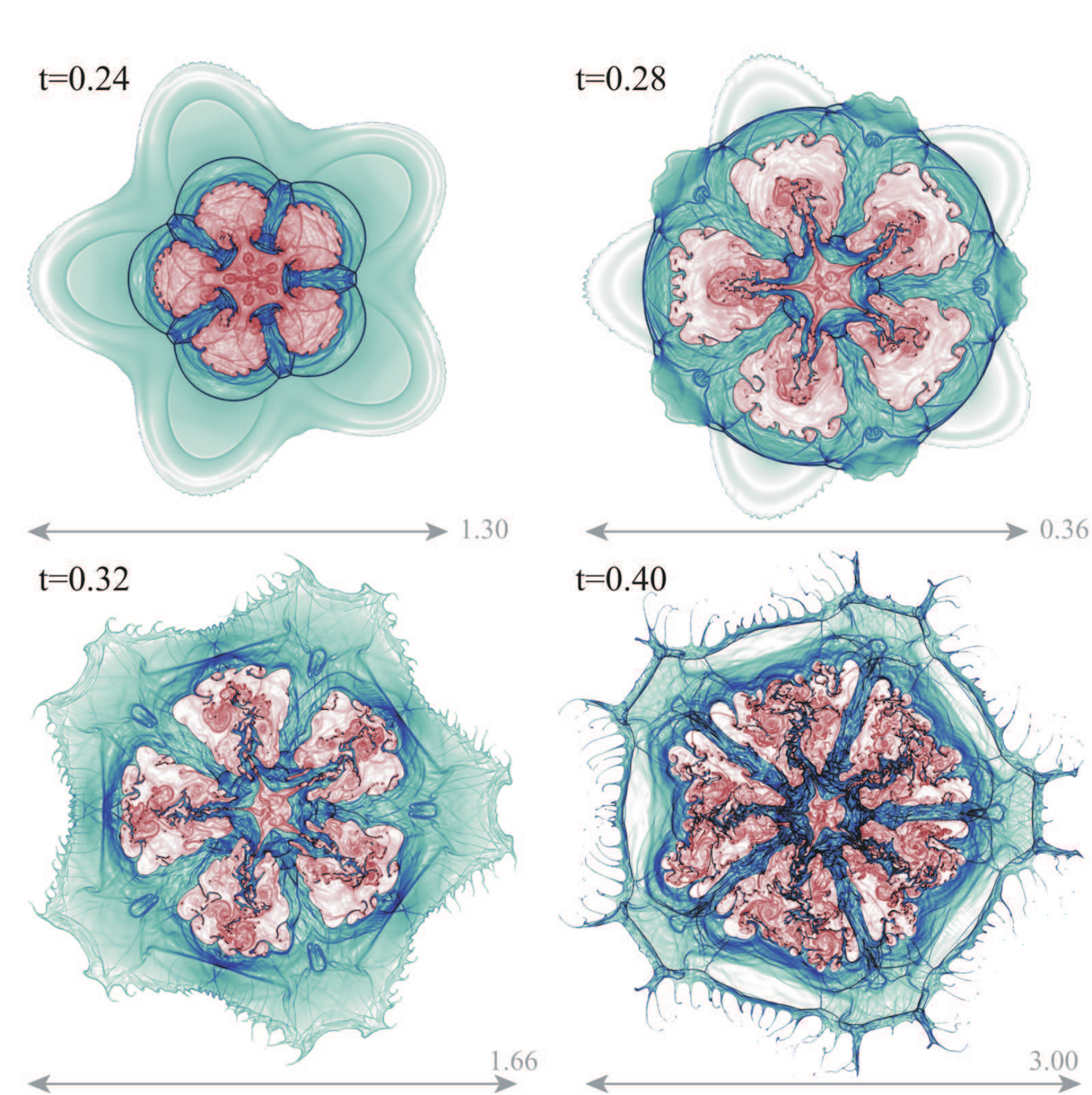}
\caption{Snapshots of density gradient after the stagnation time for 2D cylindrical ICF implosion with a high-mode perturbation.}
\label{2D_ICF_low_2}
\end{center}
\end{figure}

\begin{figure}[p]
\begin{center}
\includegraphics[height=13cm]{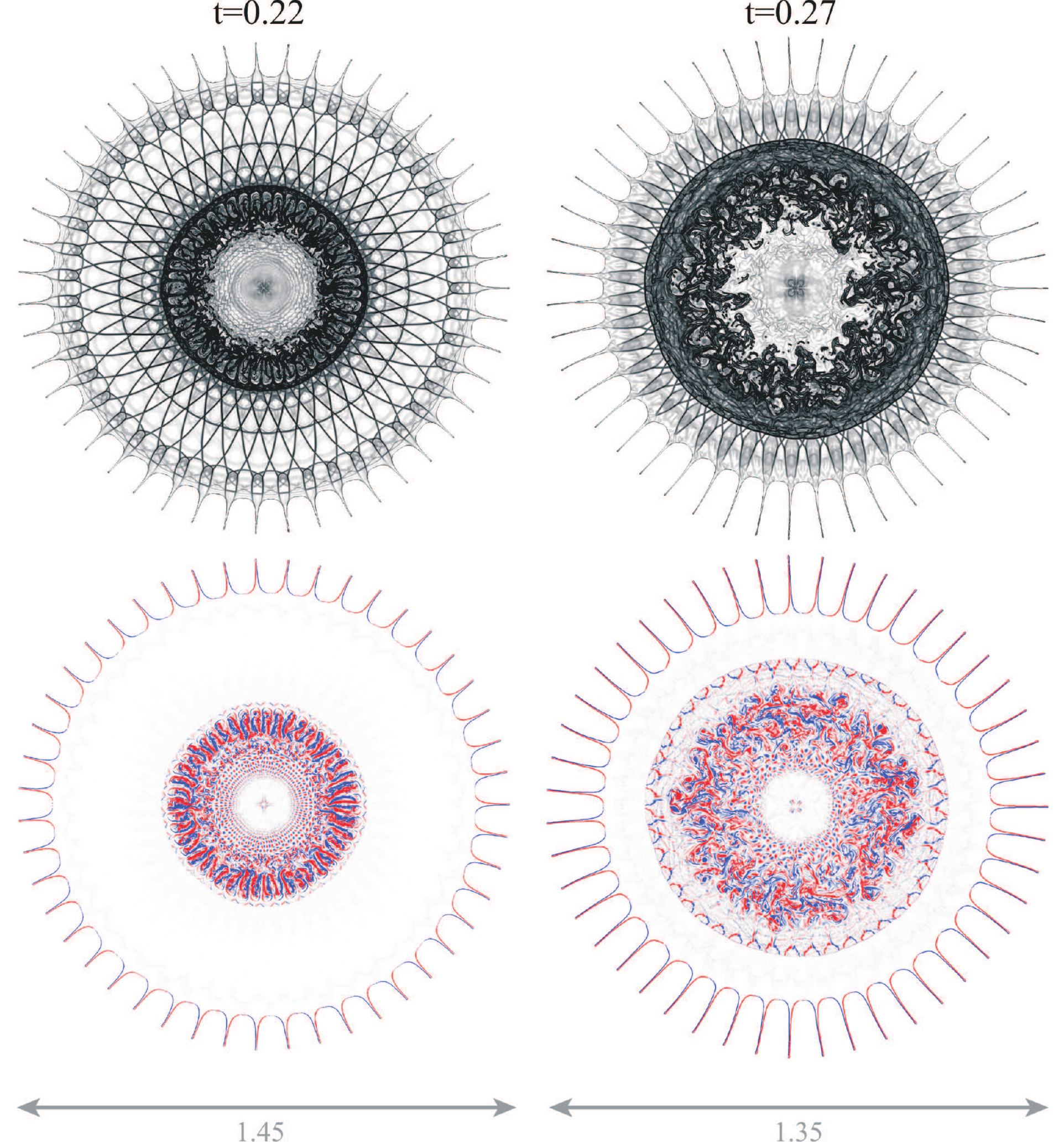}
\caption{Snapshots of density gradient and vorticity at $t=0.22$ and $0.27$ for 2D cylindrical ICF implosion with a high-mode perturbation.}
\label{2D_ICF_high}
\end{center}
\end{figure}

Figs. \ref{2D_ICF_low_1} and \ref{2D_ICF_low_2} show three snapshots of density gradient before and after the stagnation time, respectively. At the early stage, see the snapshot $t=0.12$ in Fig. \ref{2D_ICF_low_1}, the shock wave passes the inner perturbed interface and the slight distortion of the shell and light gas bubble is observed due to Richtmyer-Meshkov instability which develops linearly in time. The length scale in Fig. \ref{2D_ICF_low_1} decreases slowly. After the maximum compression time, Rayleigh-Taylor instability develops exponentially as function of time as the archived high pressure inside the light fluid bubble begins to accelerate the heavy shell. Hence the low-mode instability has grown substantially from the stagnation time, which is confirmed by rapid increase of the length scale in Fig. \ref{2D_ICF_low_2}. Some small secondary instability features near the inner interface are observed, as shown in the snapshot $t=0.28$. At a very late time $t=0.40$, the length scale plotted in Fig. \ref{2D_ICF_low_2} indicates that whole structure has significantly expanded. The instability has grown substantially and the observed features interact extensively with each other, leading to numerous small-scale mixing \cite{joggerst2014cross}. For the high-mode case, the selected two snapshots in Fig. \ref{2D_ICF_high}, $t=0.22$, $t=0.27$, are just before and after the maximum compression time. As shown in Fig. \ref{2D_ICF_high}, the outer boundary is composed of $47$ fingers and show good symmetry preservation in our simulations. The mixing zone containing most of small scales at $t=0.28$ is larger than that at $t=0.27$, see the vorticity contours in Fig. \ref{2D_ICF_high}. In both the snapshots a large number of small interface scales generate, advect and interact with each other afterwards, indicating our method is extremely robust with the aid of multi-material scale separation model.
\subsection{Shock wave interaction with a multi-material bubble (V) \label{shock_he_r22}}
\begin{figure}[p]
\begin{center}
\includegraphics[width=0.9\textwidth]{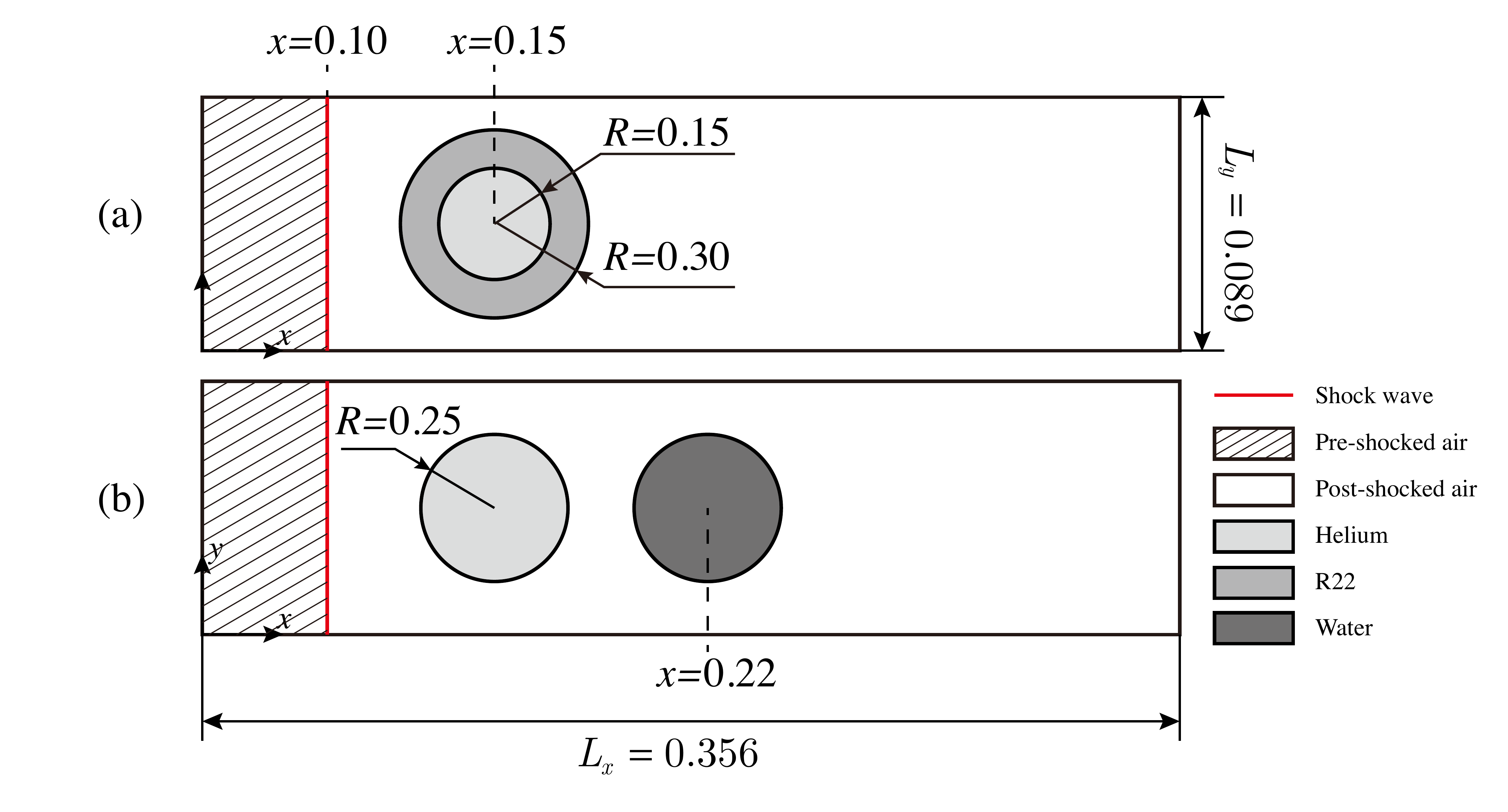}
\caption{Computational domains for 2D compressible 3-material flows: (a) case V and (b) case VI.
}
\label{setup}
\end{center}
\end{figure}

This problem, as a complex shock-accelerated inhomogeneous flow \cite{ranjan2011shock}, is a combination of 2D air-helium \cite{bagabir2001mach, hu2006conservative, hejazialhosseini2010high, han2014adaptive} and air-R22 \cite{han2014adaptive, haas1987interaction, nourgaliev2006adaptive} shock bubble interaction. A Mach $6.0$ shock wave in air will interact with a cylindrical helium bubble with a R22 shell. The computational domain and intitial conditions are shown in Fig. \ref{setup}(a) and Table. \ref{table2}, respectively. Symmetric conditions are employed at the upper and lower boundaries, while inflow and outflow conditions are prescribed at the left and right boundaries. Simulations are performed with $4 \times 1$ blocks at the coarsest level and $\ell_{max} = 7$, leading to an effective resolution of $8192 \times 2048$ at the finest level.

\begin{figure}[p]
\begin{center}
\includegraphics[width=1.0\textwidth]{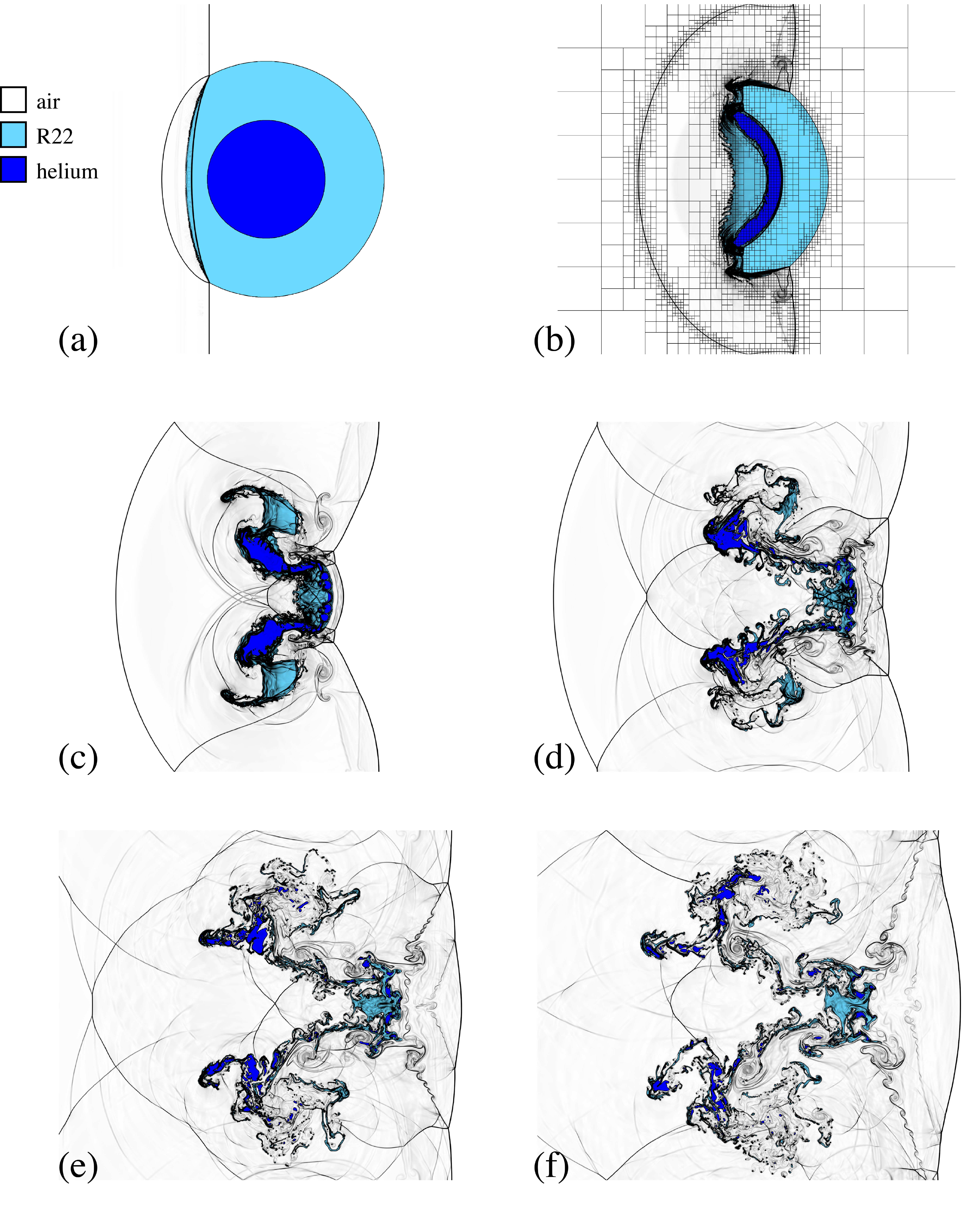}
\caption{Density gradient fields and materials distribution of case V at (a) $t=5.0 \times 10^{-3}$, (b) $t=1.0 \times 10^{-2}$, (c) $t=1.5 \times 10^{-2}$, (d) $t=2.0 \times 10^{-2}$, (e) $t=2.5 \times 10^{-2}$ and (f) $t=3.0 \times 10^{-2}$. A multi-resolution representation is outlined at $t=1.0 \times 10^{-2}$.}
\label{air-helium-r22_drho}
\end{center}
\end{figure}

\begin{figure}[p]
\begin{center}
\includegraphics[width=0.8\textwidth]{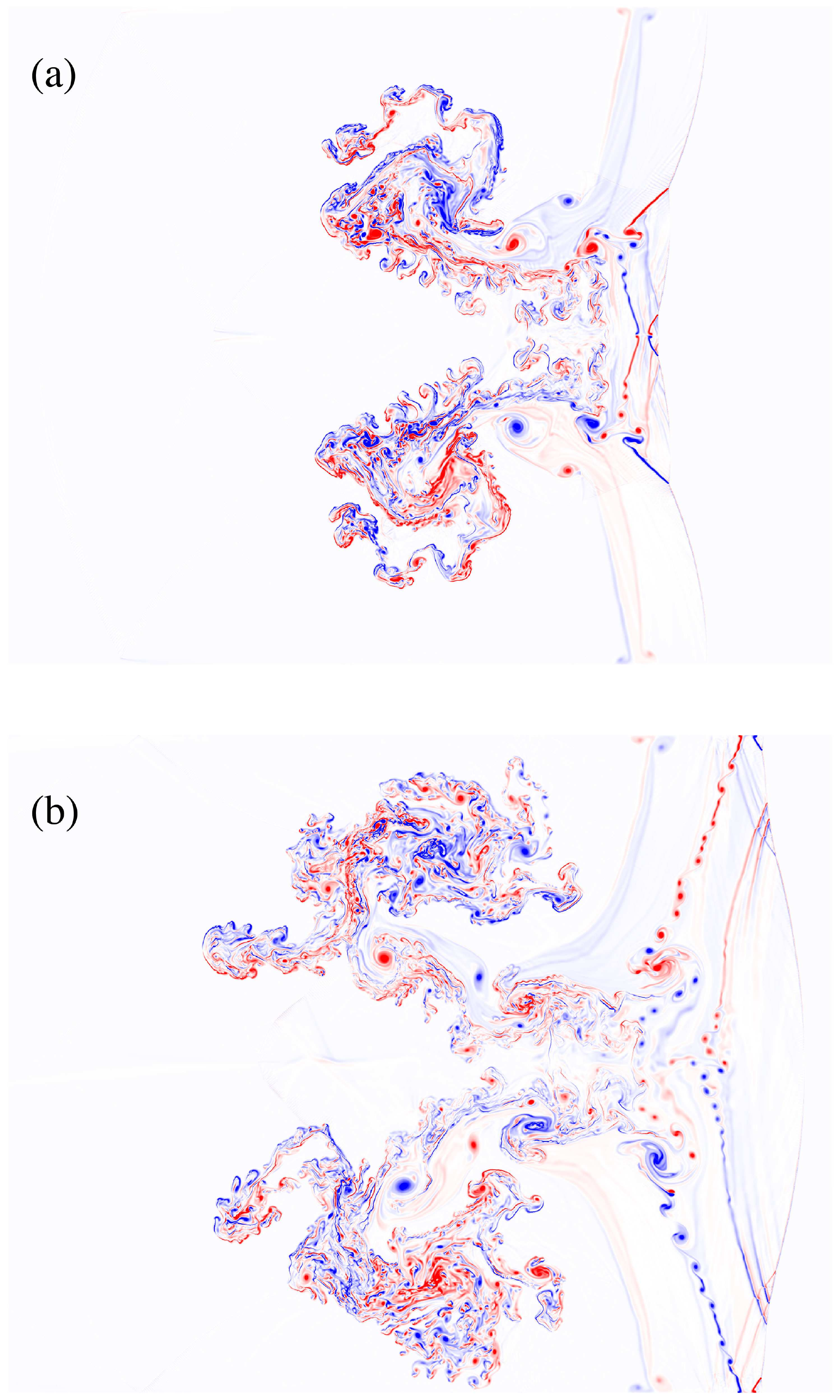}
\caption{Vorticity contours of case V at (a) $t=2.0 \times 10^{-2}$ and (b) $t=3.0 \times 10^{-2}$.}
\label{air-helium-r22_vor}
\end{center}
\end{figure}

In Fig. \ref{air-helium-r22_drho}, the density gradient fields and materials distributions at $6$ time instants illustrate the development of the shock system and the bubble deformation inside a medium with inhomogeneities of flow states. At $t=5.0 \times 10^{-3}$, the incident shock wave is refracted after crossing the upstream front of the R22 shell. As a result of the acoustic impedance mismatch $\rho_3 c_{s,3}>\rho_1 c_{s,1}$ at the air-helium interface, the transmitted shock wave has a concave curvature while the shock in the ambient air keeps planar. This convergent shock refraction pattern agrees with the numerical results in Refs. \cite{han2014adaptive, nourgaliev2006adaptive}. Meanwhile a reflected shock generates at the upstream front and then propagates upstream inside the air \cite{niederhaus2008computational, ranjan2011shock}. When the concave transmitted shock wave impacts on the R22-helium interface, the reflected rarefaction occurs and moves upstream inside the R22 shell. The acoustic impedance mismatch $\rho_2 c_{s,2}>\rho_3 c_{s,3}$ at the air-helium interface produces a convex transmitted shock inside the helium bubble. This shock propagates downstream and subsequently impacts on the downstream surface of helium bubble. At $t=1.0 \times 10^{-2}$, it moves across the interface entirely and becomes a re-transmitted shock in the R22 material, as shown in Fig. \ref{air-helium-r22_drho}(b). Correspondingly, a Mach stem, triple point, slip line and re-transmitted reflected shock (moves to the upstream of the helium bubble in Fig. \ref{air-helium-r22_drho}(b)) are produced in R22 material during this process \cite{niederhaus2008computational}. Along the vertical direction one can observe a ``fast-slow-fast-slow-fast'' type of shock speed according to definition of Zabusky and Zeng \cite{zabusky1998shock}. The deformed helium bubble has a similar shape with previous numerical results \cite{hu2006conservative, hejazialhosseini2010high, han2014adaptive}, although its incident shock is not planar here. The materials of helium and R22 are accelerated with significantly different shock speed and the global shape is extremely distorted, corresponding to ``fast-slow-fast-slow-fast'' shock wave, see Figs. \ref{air-helium-r22_drho}(c) and \ref{air-helium-r22_drho}(d). Several upstream-directed reflected shock waves with different shock strength may generate at multiple places in either R22 shell or helium bubble. These shocks will interact with each other and finally interact with the reflected shock in Fig. \ref{air-helium-r22_drho}(a), see Figs. \ref{air-helium-r22_drho}(e) and \ref{air-helium-r22_drho}(f). From $t=1.0 \times 10^{-2}$ to $t=3.0 \times 10^{-2}$, the distributions of R22 and helium are successively spread to ambient air, leading to many regions of intense mixing finally. The frequently generated small isolated R22 and helium droplets are captured in our simulation, as shown in Figs. \ref{air-helium-r22_drho}(e) and \ref{air-helium-r22_drho}(f). This complex shock system and medium inhomogeneities produce very irregular vortex structures due to baroclinical mechanism \cite{ranjan2011shock} and Kelvin-Helmholtz instabilities, as shown in Figs. \ref{air-helium-r22_vor}. A multi-resolution representation in Fig. \ref{air-helium-r22_drho}(b) confirms our mesh refinement criterion, i.e., all blocks near the interfaces and shock structures are refined to the finest level, $\ell_{max} = 7$.
\begin{table}
\centering
\resizebox{\textwidth}{!}{
\begin{threeparttable}
\caption{\label{table2} Initial conditions for 2D test cases.}
\vspace{0.5cm}
\begin{tabular}{|l|cccc|cccc|}
\hline
\multirow{4}{*}{} & \multicolumn{4}{c|}{Case V}  & \multicolumn{4}{c|}{Case VI}\\
\hline
Location\tnote{*} & Post-shocked air & Pre-shocked air & Helium bubble &  R22 shell & Post-shocked air & Pre-shocked air & Helium bubble & Water column\\
\hline $\chi$ &1	  &1	    &2	      &3	    &1	    &1	    &2	      &3\\
 $\rho$       &1.0    &5.268	&0.138	  &3.154	&1.0	&5.268	&0.138	  &1000.0\\
 $p$          &1.0	  &41.83    &1.0	  &1.0	    &1.0	&41.83	&1.0      & 1.0\\
 $u$          &5.752  &0.0      &0.0      &0.0		&5.752  &0.0	&0.0	  &0.0\\
 $\gamma$ \tnote{**}    &1.4    &1.4      &1.667	  &1.249	&1.4 	&1.4    &1.667    &7.15\\
\hline
\end{tabular}
    \begin{tablenotes}
      \small \item[*]{See Fig. \ref{setup}(a) for details.}
    \end{tablenotes}
    \begin{tablenotes}
      \small \item[**]{Tait's EOS is used for water column and idea gas EOS is for other gaseous materials.}
    \end{tablenotes}
\end{threeparttable}
}
\end{table}
\subsection{Shock wave interaction with an Helium bubble and a water column (VI) \label{shock_he_water}}

In this case, a helium bubble is initially accelerated by a Mach $6.0$ planar shock and then impacts on a water column. This can be considered as a combination of shock-helium interaction  \cite{bagabir2001mach, hu2006conservative, hejazialhosseini2010high, han2014adaptive} and shock-water interaction \cite{han2014adaptive, nourgaliev2006adaptive, chang2013direct}. The complexity arises when the shock-accelerated helium bubble impacts on water column. The computational domain and initial conditions are detailed in Fig. \ref{setup}(b) and Table. \ref{table2} while the boundary conditions are same with Sec. \ref{shock_he_r22}. We refine $4 \times 1$ blocks at the coarsest level to $\ell_{max} = 6$ to obtain an effective grid resolution of $4096 \times 1024$ at the finest level.

\begin{figure}[p]
\begin{center}
\includegraphics[width=1.0\textwidth]{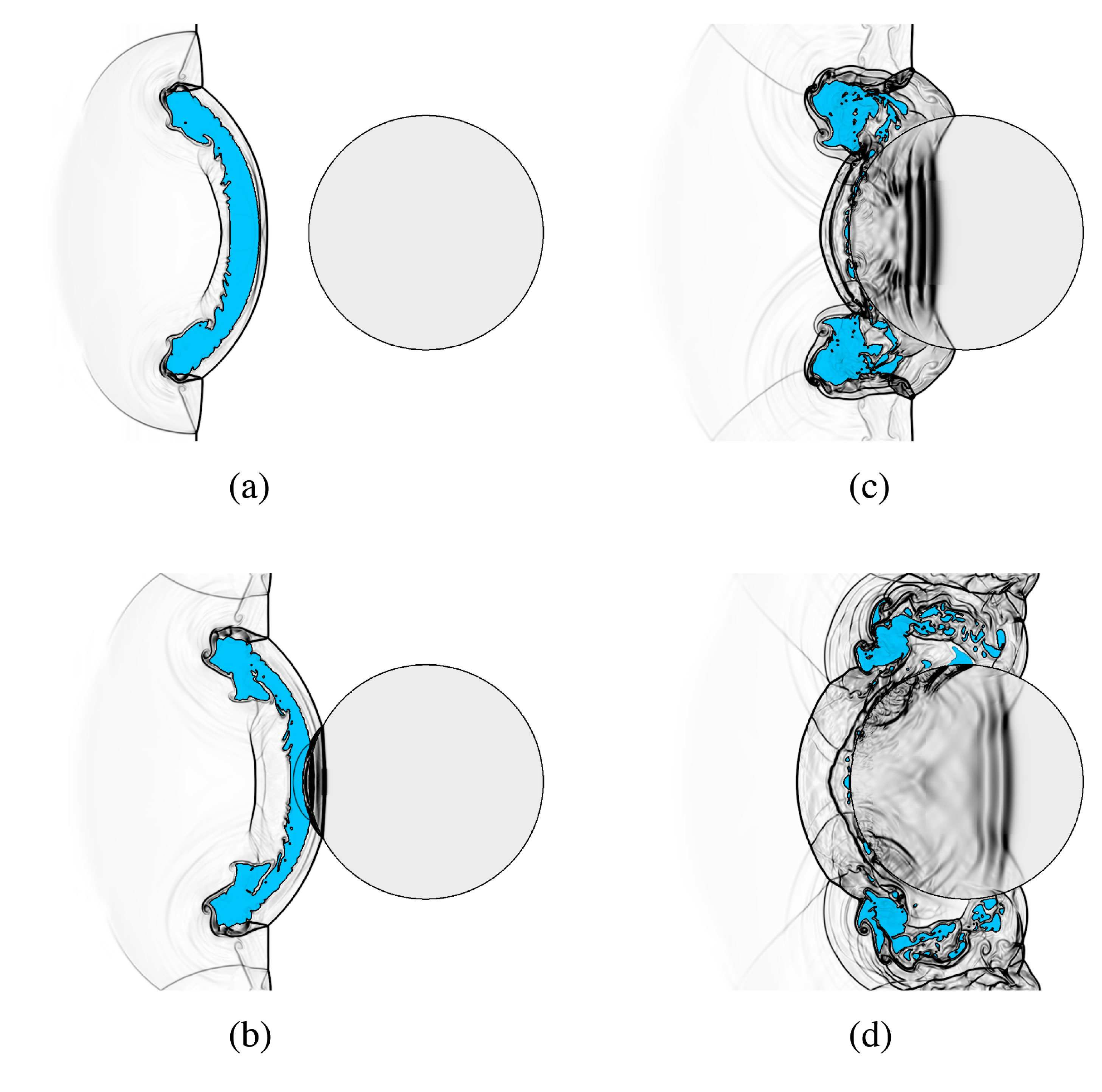}
\caption{Density gradient fields of case VI at (a) $t=1.0 \times 10^{-2}$, (b) $t=1.2 \times 10^{-2}$, (c) $t=1.5 \times 10^{-2}$ na (d) $t=1.8 \times 10^{-2}$. The distribution of helium is colored by blue to capture its interaction with water column (light gray).}
\label{air-helium-water_drho}
\end{center}
\end{figure}

We plot $4$ density gradient fields in Fig. \ref{air-helium-water_drho} at $t=1.0 \times 10^{-2}$, $1.2 \times 10^{-2}$, $1.5 \times 10^{-2}$ and $1.8 \times 10^{-2}$, with the helium colored by blue to track its deformation and distrituion after impacting on the water column. Note that the bubble deformation, the transmitted and reflected shocks, and the reflected rarefaction wave at $t=1.0 \times 10^{-2}$ are in good agreement with those in Ref. \cite{han2014adaptive}, see Fig. \ref{air-helium-water_drho}(a). The triple point, Mach stem and slip line are observed at this moment. Then this deformed helium bubble moves downstream and impacts on the water column at $t=1.2 \times 10^{-2}$. A transmitted shock and reflected shock are generated at the helium-water interface. As the speed of sound in helium is comparable with that in water, the transmitted shock in water and re-transmitted shock in the ambient air forms a bow shock wave together. Due to high stiffness of water column, the helium keeps closely touched to the water column, as shown in Fig. \ref{air-helium-water_drho}(b). From $t=1.5 \times 10^{-2}$ to $t=1.8 \times 10^{-2}$, the two roll-up regions of helium propagate across the water column, while few material of helium resides along the upstream surface of water column and small droplets are generated due to the obstacle effect of water column. Meanwhile the shock refraction pattern develops extensively and is more complex than those in previous literatures \cite{niederhaus2008computational, ranjan2011shock}.
\section{Conclusion \label{sec:conclusion}}
The proposed conservative sharp-interface method employs the multi-region level-set method, conservative finite volume method, interface interaction model and interface scale separation model to overcome typical issues in numerical simulations of compressible multi-material flows. In our numerical method, the advection of interface networks are handled by a recently developed multi-region level-set method. The conservation is strictly ensured by solving the discretized governing equation with multi-material-cell treatments and using a simple conservation correction in each multi-material-cell. The sharp interface property is achieved by an efficient reduced interface interaction model to obtain the interface condition which serves to calculate the velocity of the interface network and the exchange flux across different materials, instead of fully solving the Riemann problem inside a multi-material-cell. In addition, the robustness is enhanced by removing non-resolved interface scales with a multi-material interface scale separation model. A range
of test cases demonstrate that the proposed method is strictly conservative, highly robust, flexible and efficient for simulations of compressible multi-material flows. Although we only simulate 1D and 2D cases in this paper and the flows are restricted to invisicd, we emphasize that this method is straightforward to implement in three dimensions and have no difficulties in extending to viscous compressible flows with source terms such as surface tension and gravity by modifying the exchange flux in multi-material-cells.
\section*{Acknowledgment}
This work is supported by China Scholarship Council under No. 201306290030.
\section*{References}









\bibliographystyle{elsarticle-num}
\bibliography{aipsamp}

%
\end{document}